\documentclass[times,doublespace]{qjrms4}
\usepackage[colorlinks,bookmarksopen,bookmarksnumbered,citecolor=red,urlcolor=red]{hyperref}
\usepackage{algorithm,algpseudocode}

\usepackage{subcaption}
\usepackage[toc,page]{appendix}
\usepackage{natbib}  
\usepackage{titlesec}
\usepackage{soul}
\newcommand\BibTeX{{\rmfamily B\kern-.05em \textsc{i\kern-.025em b}\kern-.08em
T\kern-.1667em\lower.7ex\hbox{E}\kern-.125emX}}

\usepackage{moreverb}

\def\volumeyear{2017}
\newcommand{\bx} { \mathbf{x}} 
\newcommand{\by} {\mathbf{y}}
\newcommand{\bz} {\mathbf{z}}
\newcommand{\bw} {\mathbf{w}}

\newcommand{\bidx} {\mathbf{idx}}
\newcommand{\bff} {\mathbf{f}}
\newcommand{\bu} {\mathbf{u}}
\newcommand{\bX} { \mathbf{X} }
\newcommand{\bA} { \mathbf{A} }
\newcommand{\bD} {\mathbf{D} }
\newcommand{\bc} {\mathbf{c} }
\newcommand{\bd} {\mathbf{d} }
\newcommand{\bH} {\mathbf{H} }
\newcommand{\bI} { \mathbf{I} }
\newcommand{\bJ} { \mathbf{J} }
\newcommand{\bK} { \mathbf{K} }

\newcommand{\bL} { \mathbf{L} }
\newcommand{\bP} { \mathbf{P} }
\newcommand{\bR} {\mathbf{R} }
\newcommand{\bS} {\mathbf{S} }
\newcommand{\bQ} { \mathbf{Q} }
\newcommand{\bT} { \mathbf{T} }
\newcommand{\bU} { \mathbf{U} }
\newcommand{\bV} { \mathbf{V} }
\newcommand{\bW}{\mathbf{W}}
\newcommand{\bbeta}{\boldsymbol{\beta}}
\newcommand{\bgamma}{\boldsymbol{\gamma}}
\newcommand{\bepsilon}{\boldsymbol{\epsilon}}
\newcommand{\bmu}{\boldsymbol{\mu}}
\newcommand{\bnu}{\boldsymbol{\nu}}
\newcommand{\bxi}{\boldsymbol{\xi}}

\newcommand{\pj}[1]{{\color{red}#1}}

\begin{document}
\graphicspath{{figures/}}

\runningheads{Peter~Jan~van~Leeuwen,~et~al.}{Particle Filters for Applications in Geosciences}

\title{Particle filters for high-dimensional geoscience applications: A Review \footnotemark[2]}

\author{Peter Jan van Leeuwen\affil{a,b}\corrauth, Hans R. K\"unsch\affil{c},Lars Nerger\affil{d}, Roland Potthast\affil{e}, Sebastian Reich\affil{f}}

\address{\affilnum{a}Department of Meteorology and National Centre for Earth Observation, University of Reading, Reading, UK\\
                \affilnum{b}Department of Atmospheric Sciences, Colorado State University, USA\\
                 \affilnum{c}Seminar f{\"u}r Statistik, ETH Zurich, Swiszerland\\
               \affilnum{d}Alfred Wegener Institute Helmholtz Center for Polar and Marine Research, Bremerhaven, Germany\\
               \affilnum{e} Deutscher Wetterdienst, Offenbach, Germany\\
                \affilnum{f}Institut f{\"u}r Mathematik, Universitaet Potsdam,  Potsdam, Germany
               }

\corraddr{Peter Jan van Leeuwen, Department of Meteorology, University of Reading, Reading RG6 6BB, UK. E-mail: p.j.vanleeuwen@reading.ac.uk}

\begin{abstract}

Particle filters contain the promise of fully nonlinear data assimilation.
They have been applied in numerous science areas, including the geosciences, 
but their application to high-dimensional geoscience systems
has been limited due to their inefficiency in high-dimensional systems
in standard settings. However, huge progress has been made, and this limitation
is disappearing fast due to recent developments in proposal densities, 
the use of ideas from (optimal) transportation,
the use of localisation and intelligent adaptive resampling strategies. 
Furthermore, powerful hybrids between particle filters
and ensemble Kalman filters and variational methods have been developed. 
We present a state of the art discussion of present efforts of developing particle filters
for high-dimensional nonlinear geoscience state-estimation problems with an 
emphasis on atmospheric and oceanic applications,
including many new ideas, derivations, and unifications, highlighting hidden connections,
including pseudo code,
and generating a valuable tool and guide for the community.
Initial experiments show that particle filters can be competitive with present-day methods for 
numerical weather prediction suggesting that they will become mainstream soon.

\end{abstract}


\maketitle

\footnotetext[2]{Please ensure that you use the most up to date class file, available from the QJRMS Home Page at \\
\href{http://onlinelibrary.wiley.com/journal/10.1002/(ISSN)1477-870X}{\tiny\texttt{http://onlinelibrary.wiley.com/journal/10.1002/(ISSN)1477-870X}}%
}

\section{Introduction}
\label{sec:intro}

Data assimilation for geoscience applications, such as weather or ocean prediction, is a slowly maturing field.
Even the linear data-assimilation problem cannot be solved adequately because of the size of the problem.
Typically, global-scale numerical weather prediction needs estimation of over $10^9$ state variables, assimilating 
over $10^7$ observations every 6-12 hours. 
Existing methods like 4DVar do not provide accurate uncertainty estimates and need efficient pre-conditioners, 
while Ensemble Kalman Filters heavily
rely on somewhat ad-hoc fixes like localisation and inflation to find accurate estimates.
Hybrids of variational and ensemble Kalman filter methods are a step forward, although localisation
and inflation are still needed in realistic applications. An extra complication is localisation 
over time needed in ensemble smoothers like the Ensemble Kalman Smoother and 4DEnsVar
when the fluid flow is strong: what is local at observation time is not necessary local
at the start of the assimilation window because the observation influence is advected with the flow.
Furthermore, the recent surge of papers on accurate treatment of observation errors shows that a long way
is still ahead of us to solve even the (close to) linear data-assimilation problem.

Although these problems are formidable, another difficulty arises from the fact that the problem is typically nonlinear,
and, with increasing model resolution and more complex observation operators,  increasingly so.
Both variational and Kalman-filter-like methods have difficulty handling nonlinear problems.
Variational methods can easily fail when the cost function is multimodal, and are hampered by the assumption
that the prior probability density function (pdf) of the state is assumed to be Gaussian. 
Ensemble Kalman filters make the explicit assumption that the prior pdf and the likelihood of the observations as function of the state
are Gaussian, or, somewhat equivalently, assume that the analysis is a linear combination of prior state and 
observations.
Both methods have been shown to fail for nonlinear data-assimilation problems in low-dimensional systems, 
and both have been reported to have serious difficulties in numerical weather prediction at the convective scale
where the model resolution is only a few km. Particle filters hold the promise of fully nonlinear data assimilation 
without any assumption on prior or likelihood,
and recent text books like \cite{reichcotter15}, \cite{Nakamura2015}, and \cite{VanLeeuwen15} 
provide useful introductions to data-assimilation in general, and particle filters in particular.

Other fully nonlinear data-assimilation methods are Markov-Chain Monte-Carlo methods that draw directly from the posterior in a sequential way, so one sample
after the other, after a burn-in period, see e.g. \cite{Robert04}, or \cite{VanLeeuwen15} for a geophysics-friendly 
introduction. The samples are correlated, often $100\% $ when the new sample is not accepted, making them
very inefficient in high-dimensional systems. This is why we concentrate on particle filtering here.

The standard or bootstrap particle filter can be described as follows.
The starting point is an ensemble of size $N$ of model states $\bx_i^n \in \Re^{N_x}$, called particles, that represent the prior 
probability density function (pdf) $p(\bx^n)$, as:
\begin{equation}
\label{eq:prior1}
p(\bx^n) \approx \sum_{i=1}^N \frac{1}{N} \delta(\bx^n-\bx_i^n)
\end{equation}
Between observations, each of these particles is propagated forward from time $n-1$ to time $n$ with the 
typically nonlinear model equations
\begin{equation}
\label{eq:model}
\bx^n = f(\bx^{n-1}) + \bbeta^n
\end{equation}
in which $f(..)$ denotes the deterministic model, and $\bbeta^n$ is a random forcing representing missing physics,
discretisation errors, etc. In this paper we assume this model noise to be additive, but one could also
consider multiplicative noise in which $\bbeta^n$ is a function of the state of the system.
We assume that the pdf from which the $\bbeta^n$ are drawn is known; typically a Gaussian $N(0,\bQ)$.

At observation times the true system is observed via:
\begin{equation}
\by^n = \bH(\bx^n_{true}) + \bepsilon^n
\end{equation}
in which the observation errors $\bepsilon^n$ are random vectors representing measurement errors and possibly
representation errors. Again we assume that these errors have known characteristics, often Gaussian, so e.g. $\bepsilon^n \sim N(0,\bR)$.
These observations $\by^n \in \Re^{N_y}$ are assimilated by multiplying the prior pdf above with the likelihood of each possible state,
i.e.\ the probability density  $p(\by^n|\bx^n)$ of the observation vector given each possible model state,
following Bayes Theorem:
\begin{equation}
\label{eq:bayes}
p(\bx^n|\by^n) = \frac{p(\by^n|\bx^n)}{p(\by^n)}p(\bx^n)
\end{equation}
in which $p(\bx^n|\by^n)$ is the posterior pdf, the holy grail of data assimilation.
\pj{To avoid confusion, it is good to realise that the true state is not a random variable when we apply
Bayes Theorem. It is a realisation of a process, which could be random or deterministic, from which we then take 
noisy observations. Instead, Bayes Theorem is a statement of what we think the true state might be.
Since the pdf of the $\bepsilon^n$ is known and Bayes Theorem is a statement for each possible
state $\bx^n$ to be the true state, $p(\by^n|\bx^n)$ is the pdf of $\by^n$ given that the true state vector would be $\bx^n$. 
In general, since for a given state $\bx^n$ the observation $\by^n$ is equal to the observation error $\bepsilon$ shifted by $\bH(\bx^n)$, we find (see e.g. \cite{VanLeeuwen2015c})}:
\begin{equation}
p(\by^n|\bx^n) = p_{\bepsilon}(\by^n-\bH(\bx^n))
\end{equation}

\begin{figure}[h]
	\centering\includegraphics[width=1\linewidth]{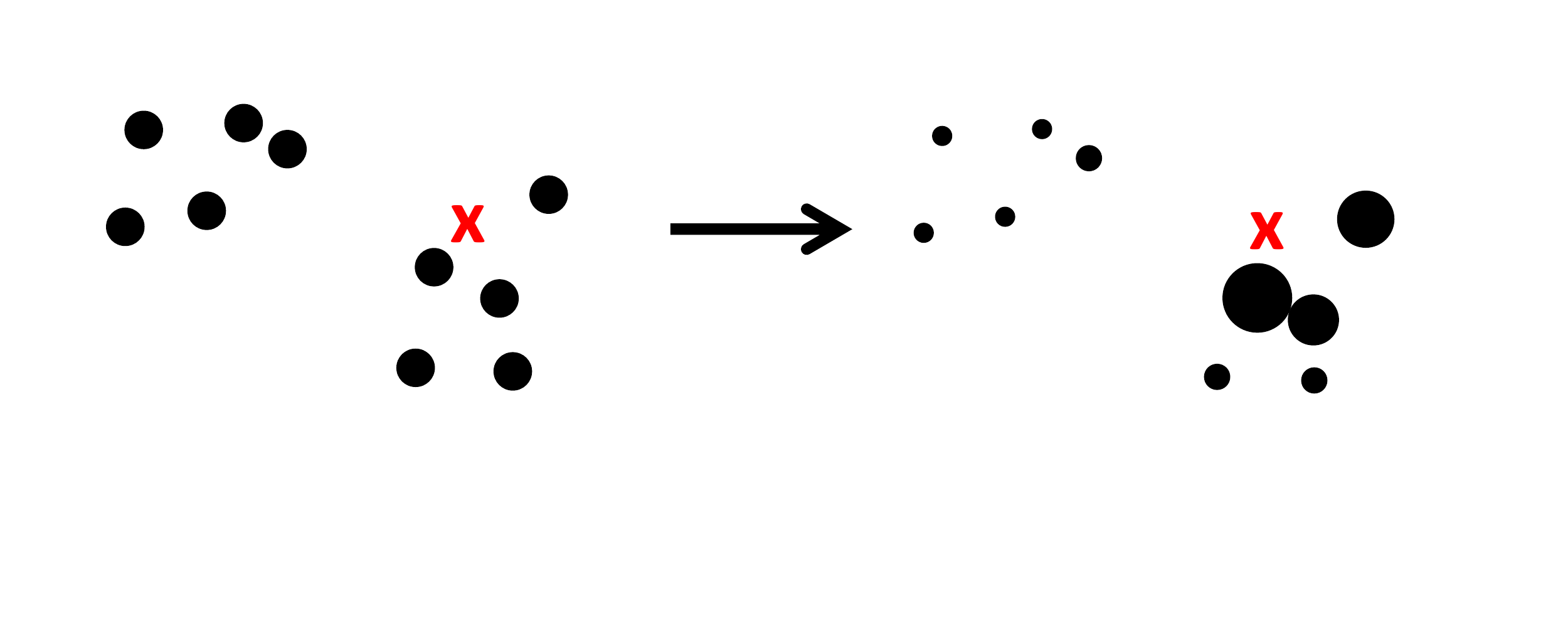}   
	\caption{The standard particle filter. Left: the prior particles (dots), with one observation,
	denoted with the red cross. Right: the posterior particles, the larger the dot the larger its weight.
	Note that the particles don't move in state space, they are just reweighted.} 
	\label{fig:Standard-PF}
\end{figure}

If we insert our particle representation of the prior into this theorem we find:
\begin{equation}
p(\bx^n|\by^n) \approx \sum_{i=1}^N w_i \delta(\bx^n-\bx_i^n)
\end{equation}
in which the particle weights $w_i$ are given by:
\begin{equation}
w_i^n = \frac{p(\by^n|\bx_i^n)}{Np(\by^n)} = \frac{p(\by^n|\bx_i^n)}{N\int p(\by^n|\bx^n)p(\bx^n)\;d\bx^n} \approx \frac{p(\by^n|\bx_i^n)}{\sum_j p(\by^n|\bx_j^n)}
\label{eq:weights}
\end{equation}
Since all terms are known explicitly we can just calculate this as a number.
The self-normalisation in the last part of (\ref{eq:weights}) is consistent with the notion that for a proper representation of
a pdf the sum of the weights should be equal to one, so that the integral over the whole state
space of the particle representation of the pdf is equal to one.
Figure \ref{fig:Standard-PF} depicts the working of this filter.

Propagating the particles $\bx_i^n$ to the next observation time $n +1$ gives a weighted 
representation of the prior at time $n + 1$. Assimilating the observation at time $n + 1$ by Bayes Theorem
leads to a modification of the weights (see e.g. \cite{Doucet2001} or \cite{VanLeeuwen2009}):
\begin{equation}
w_i^{n+1} = w_i^n \frac{p(\by^{n+1}|\bx_i^{n+1})}{\sum_j p(\by^{n+1}|\bx_j^{n+1})}
\end{equation}
Even in low-dimensional applications, the variation of the weights
increases with the number of assimilation steps. Eventually one
particle has a much higher weight than all the others. To prevent
this, resampling can be used before propagation to obtain equally
weighted particles. This duplicates high-weight particles and abandons low-weight particles.
After resampling, some of the particles have
identical values, but if the model contains a stochastic component
and independent random forcings are used for different particles,
diversity is restored. See e.g. \cite{Doucet2001} or \cite{VanLeeuwen2009} for details.
Algorithm \ref{alg:SPF} illustrates the steps.

\begin{algorithm}[!ht]
	\caption{Standard Particle Filter \label{alg:SPF}}
	\begin{algorithmic}[0]
            \For {$i=1,..,N$}
	    \State $ w_i \gets p(\by | \bx_i^n) $
	   \EndFor
	   \State $ \bw \gets \bw / \bw^T \mathbf{1} $
	   \State Resample
	\end{algorithmic}
\end{algorithm}
A simple resampling scheme using only one draw from a uniform distribution {\em U} is presented in Algorithm \ref{alg:Resample}.
\begin{algorithm}[!ht]
	\caption{Simple Resampling Scheme \label{alg:Resample}}
	\begin{algorithmic}[0]
           \State $\hat{w}_1 \gets w_1$
	 \For{$j =2, ..,N$} 
		\State $\hat{w}_j = \sum_{i=1}^j w_j$
	\EndFor
	\State $u \sim U[0,1/N]$ 
	\State $m \gets 1$
	\For{$j =2, ..,N$} 
		\While{$u > \hat{w}_m$}
			\State $m \gets m+1$
			\State $ \bx_m^{new} = \bx_j $
		\EndWhile
		\State $u \gets u+1/N$
	\EndFor
	\end{algorithmic}
\end{algorithm}

In high-dimensional problems the weights vary enormously even at one observation time, and typically one particle 
obtains a much higher weight than all the others.
\cite{Snyder2008, Snyder2015} have shown that the number of particles needed
to avoid weights collapse, in which one particle gets weight 1 and the rest weights very close to zero, 
has to grow exponentially with the dimension of the observations $\by$ for a large class of particle filters.
If the weights collapse, all particles are identical after resampling, and all diversity is lost. From this discussion it becomes clear that for particle filters to work we need to ensure that
their weights remain similar.

In this review we will discuss four basic ways to make progress on this fundamental problem
of weight degeneracy.
In the first one, we explore the so-called proposal-density freedom to 
steer particles through state space such that they obtain very similar weights, see e.g. \cite{Doucet2001}. 
As pointed out by e.g.  \cite{Snyder2008} there 
are fundamental problems when applying these techniques to the high-dimensional 
geoscience applications. We will examine the issue in detail and discuss so-called equal-weight particle filters, 
which point towards new ways to formulate and attack the degeneracy problem. 

The second approach transform the prior particles into particles from 
the posterior, either in one go, or via a more smooth transformation process, see \cite{reich13}. 
While the one-step approaches can be shown to fail in high-dimensional settings, they do lend themselves 
very naturally to localisation. The more smooth multi-step transition variants
seem to be able to avoid the degeneracy problem without localisation, and are
an interesting new development.

The third, more straightforward from the geoscience experience, approach is to introduce localisation in particle filters.
While initial implementations were discouraging (e.g. Van Leeuwen, 2009), new formulations have shown remarkable successes,
such that localised particle filters are now tested in global operational 
numerical weather prediction systems \citep[e.g.][]{potthast17_1}.

The fourth approach is to abandon the idea of using pure particle filters and 
combine them with Ensemble Kalman Filters. This should not be confused with 
using Ensemble Kalman Filters in proposal densities. Several variants exist, 
such as second-order exact filters, in which only the first two moments are 
estimated, sequential versions in which first an EnKF is used and the 
posterior EnKF ensemble is used as input for the particle filter, or vice versa, 
and combinations in which localised weights are calculated and dependent on the effective ensemble
size a full particle filter, an EnKF, or a combination of both is used. 

These four variants form the basis of the following four chapters. Each chapter contains a 
critical discussion of the approximations and remaining major issues.  It should be noted that the
pseudo code provided does not give the most efficient implementation of the different
particle filters, but is rather an illustration of the computational steps involved. Efficient pseudo code for some of the more complex schemes can be found in \cite{Vetra2018}.
The paper is closed with
a concluding section and an outlook of what possible next steps could be.

\section{Proposal density particle filters}
\label{sec:proposal_density_pfs}

Ideally we draw independent samples directly from the posterior pdf because the samples would all have equal weight
automatically. This can only be done, however, when the shape of the posterior pdf is known and when it is easy to
draw from the posterior. An example of this is a Gaussian prior combined with a linear Gaussian likelihood. Under these assumptions the
posterior is also Gaussian and the mean and covariance can be calculated directly from the prior using the Kalman update
equations. Ensemble Kalman filters make use of this result and draw directly from that pdf, which is why all posterior particles have equal
weights in an Ensemble Kalman Filter.

The standard particle filter draws particles from the prior. These then have to be modified to become particles 
of the posterior via the weighting with the likelihood. This is a general
procedure in statistics called importance sampling: one draws from
an approximation of the pdf one is interested in, and corrects for this 
via so-called importance weights. 

In the introduction we argued that drawing from the prior leads to weights that vary too much: typically, 
in high-dimensional problems with numerous independent observations one particle gets weight 1, and
all other particles have a weight very close to zero.
However, we could explore the idea of importance sampling on the transition from one time to the next.
When the numerical model is not deterministic but stochastic we have the freedom 
to change the model equations to move the particles to those parts of state space where we want them to be,
for instance closer to the observations. 

Mathematically this works as follows. Assume we have observations at time
$n$, so Bayes Theorem at time $n$ is given by (\ref{eq:bayes}).
If the model is stochastic, we can write the prior as
\begin{equation}
\label{eq:prior}
p(\bx^n) = \int p(\bx^n|\bx^{n-1}) p(\bx^{n-1})\;d\bx^{n-1} 
\end{equation}
where $p(\bx^n|\bx^{n-1})$ is the transition density, the pdf of the state at
time $n$ when the state at time $n-1$ is known. For
instance, if the model error is additive and the model equation is given by (\ref{eq:model}),
it holds that 
\begin{equation}
p(\bx^n|\bx^{n-1}) = p_{\bbeta}\left(\bx^n-f(\bx^{n-1})\right).
\end{equation}
Often the model errors are assumed to be Gaussian
$\bbeta \sim N(0,\bQ)$, and we find
\begin{equation}
\label{eq:model-Gauss}
p(\bx^n|\bx^{n-1}) = N(f(\bx^{n-1}),\bQ).
\end{equation}
but the method is more general than that.

Assume now that at time $n-1$ we have a set of weighted particles as in (\ref{eq:prior1}),
but with weights $w_i^{n-1}$ instead of $1/N$.
We can evaluate the expression \eqref{eq:prior} for the prior as
a weighted mixture of transition densities
\begin{equation}
\label{eq:mix-prior}
p(\bx^n) \approx \sum_{i=1}^N w^{n-1}_i p(\bx^n|\bx_i^{n-1}) 
\end{equation}
In the following we neglect the approximation error at time $n-1$
and assume that \eqref{eq:mix-prior} is exact. 
This is not necessarily a good approximation, especially when the number of particles 
is small. On the other hand, it is consistent with the particle filter approximation in the first place,
and one of the few things one can do.
By Bayes formula
\eqref{eq:bayes}, the posterior can then be written as:
\begin{equation}
\label{eq:mix-posterior}
p(\bx^n|\by^n) \approx \sum_{i=1}^N w^{n-1}_i \frac{p(\by^n|\bx^n)}{p(\by^n)}p(\bx^n|\bx_i^{n-1}) 
\end{equation}

In the standard particle filter one makes one draw from
$p(\bx^n|\bx_i^{n-1})$ for each $i$, 
and we know that this leads to ensemble collapse for high-dimensional systems.
However, now the prior particles at time $n$ are allowed to arise from
following a different model equation. This works as follows. 
We can multiply and divide equations \eqref{eq:mix-prior}
and \eqref{eq:mix-posterior} by a so-called proposal density
$q(\bx^n|\bx^{n-1},\by^n)$, leading to:
\begin{equation}
  \label{eq:mix-prior2}
p(\bx^n) \approx \sum_{i=1}^N w^{n-1}_i \frac{p(\bx^n|\bx_i^{n-1})}{q(\bx^n|\bx_i^{n-1},\by^n)} q(\bx^n|\bx_i^{n-1},\by^n)
 \end{equation} 
and
\begin{equation}
\label{eq:mix-posterior2}
p(\bx^n|\by^n) \approx \sum_{i=1}^N w^{n-1}_i \frac{p(\by^n|\bx^n)}{p(\by^n)}\frac{p(\bx^n|\bx_i^{n-1})}{q(\bx^n|\bx_i^{n-1},\by^n)} q(\bx^n|\bx_i^{n-1},\by^n)
\end{equation}
where $q(\bx^n|\bx_i^{n-1},\by^n)$ should be non-zero whenever $p(\bx^n|\bx_i^{n-1})$ is.
This step is completely general. 

\begin{figure}[h]
	\centering\includegraphics[width=1\linewidth]{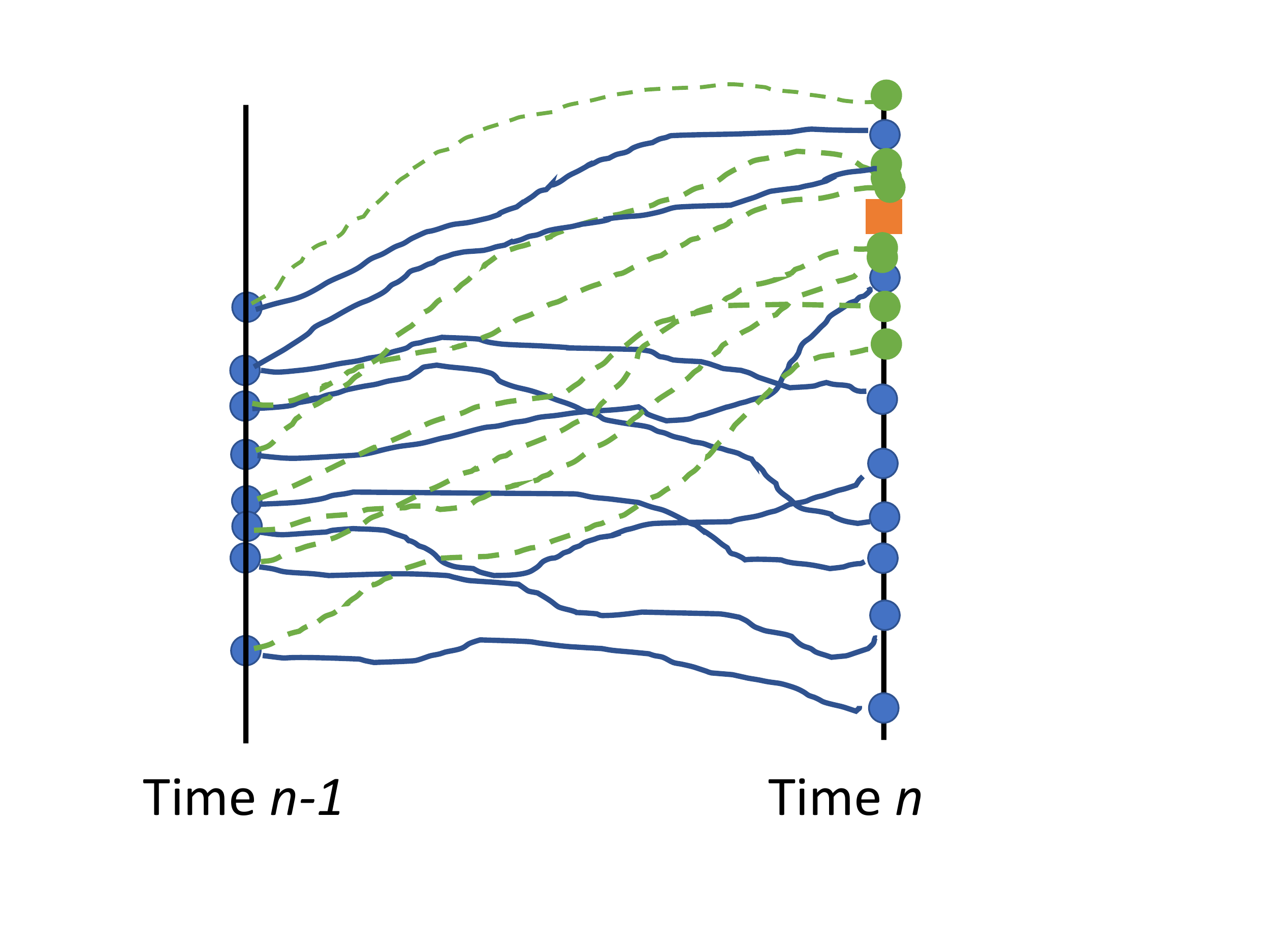}   
	\caption{The proposal density. At time $n-1$ we have a set of particles denoted by the
	filled circles. When we use the original model, they are propagated along the
	blue lines to time $n$. Because their distance to the observation (the box) varies 
	significantly, so will their weights. When a proposed model is used the particles at time $n-1$
	propagate along the green dashed lines and end up much closer to the observations. This leads
	to much more similar likelihood weights. However, because we have changed the model equations 
	the particles now also have proposal weights.} 
	\label{fig:proposal}
\end{figure}

Now realise that drawing from $p(\bx^n|\bx_i^{n-1})$
corresponds to running the original stochastic model.
We could instead draw from $q(\bx^n|\bx_i^{n-1},\by^n)$,
which would correspond to a model equation from our choosing.
Figure \ref{fig:proposal} illustrates the basic idea.

For instance when the original model is given by
\eqref{eq:model}, we can use 
\begin{equation}
\bx^{n} = g(\bx^{n-1},\by^n) + \hat{\bbeta}^n
\end{equation}
in which $g(.,.)$ is now the deterministic part and $\hat{\bbeta}^n$ is the
stochastic part. These can be freely chosen, and examples of these will be given below.
Note that we allowed $g(..)$ to depend on the observations at the future
time. This means that we generate the prior particles at time $n$ by making
one draw from $q(\bx^n|\bx^{n-1}_i,\by^n)$ for each $i$ where
\begin{equation}
q(\bx^n|\bx^{n-1},\by^n) =p_{\hat{\bbeta}}\left(\bx^n-g(\bx^{n-1},\by^n)\right)
\end{equation}
In general, we draw the particles at time $n$ from the 
alternative model $q(\bx^n|\bx^{n-1},\by^n)$ and account for this by
changing the weights of the particles. Equations \eqref{eq:mix-prior2}
and \eqref{eq:mix-posterior2} can be written as
  \begin{equation}
    p(\bx^n) = \sum_{i=1}^N \hat{w}^{n-1}_i q(\bx^n|\bx_i^{n-1},\by^n)
  \end{equation}
and
\begin{equation}
  p(\bx^n|\by^n) = \sum_{i=1}^N \hat{w}^n_i q(\bx^n|\bx_i^{n-1},\by^n)
\end{equation}
where the weights are given by:
\begin{equation}
  \hat{w}^{n-1}_i  \propto w^{n-1}_i \frac{p(\bx_i^n|\bx_i^{n-1})}{q(\bx_i^n|\bx_i^{n-1},\by^n)} .
\end{equation}
and
\begin{equation}
   \hat{w}^n_i  \propto \hat{w}^{n-1}_i \frac{p(\by^n|\bx_i^n)}{p(\by^n)}
    \propto w^{n-1}_i p(\by^n|\bx_i^n) \frac{p(\bx_i^n|\bx_i^{n-1})}{q(\bx_i^n|\bx_i^{n-1},\by^n)}.
  \end{equation}
Here the coefficients of
proportionality ensure that the weights sum to 1. 
In a reinterpretation of these equations, if $\bx_i^n$ is drawn from the
alternative model $q(\bx^n|\bx^{n-1}_i,\by^n)$ we can also write
\begin{equation}
p(\bx^n) \approx \sum_{i=1}^N \hat{w}^{n-1}_i \delta(\bx^n-\bx_i^n)
\end{equation}
and
\begin{equation}
p(\bx^n|\by^n) \approx \sum_{i=1}^N \hat{w}^{n}_i \delta(\bx^n-\bx_i^n).
\end{equation}

\begin{figure}[h]
	\centering\includegraphics[width=1\linewidth]{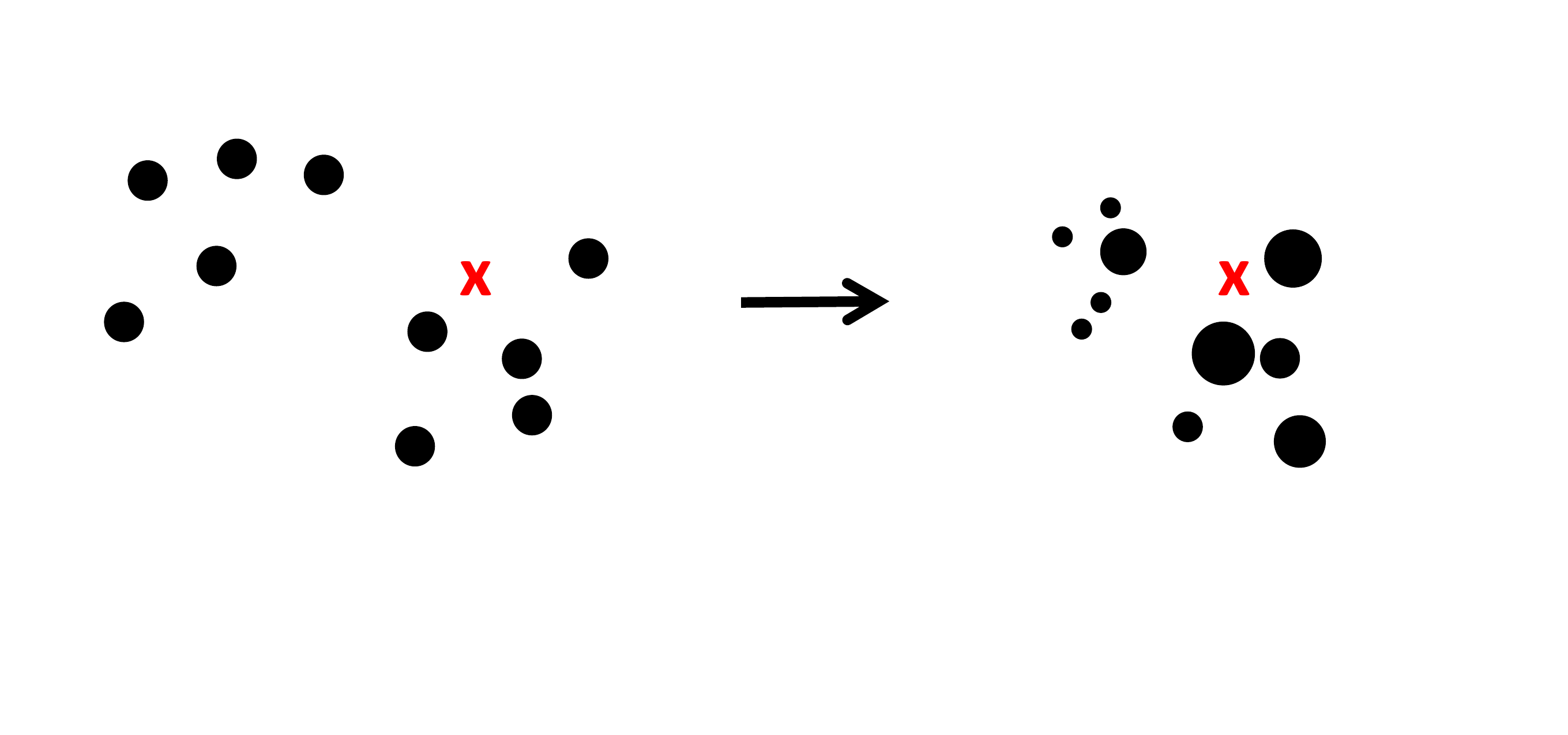}   
	\caption{The typical proposal-density particle filter. Left: the prior particles at time $n-1$ (dots), with one observation,
	denoted with the red cross. Right: the posterior particles at time $n$, the larger the dot the larger its weight.
	Note that the particles do move in state space compared to a pure model propagation over one time step, 
	and their weight contains contributions from the likelihood and from that movement.} 
	\label{fig:Optimal-PF}
\end{figure}

We see that the weights now contain two factors, the likelihood weight, which also appears in the
standard particle filter, and a proposal weight. These two weights have opposing effects.
If we use a proposal density that strongly pushes the model towards the observations, the likelihood weight
will be large because the difference between observations and model states becomes smaller, but the proposal weight 
becomes smaller because the model is pushed away from where it wants to go, so $p(\bx^n|\bx^{n-1}_i)$
will be small. On the other hand, a weak pushing towards the observations keeps the proposal weight high,
but leads to a small likelihood weight. This suggests that there is an optimum weight related to an optimal
position $\bx_i^n$ for each particle as function of its position at time $n-1$. This will be explored
in equal-weight formulations of the particle filter.
Figure \ref{fig:Optimal-PF} shows how typical proposal-density particle filters work. Equal-weight particle filters are discussed later.

\subsection{A simple relaxation scheme}
To illustrate the idea of a proposal density we consider the following simple example.
We could add a relaxation or nudging term
to the original equation to steer the particles towards the observations and
make their weights more similar, as pioneered by \cite{VanLeeuwen2010} for geoscience 
applications. The model equation is written as:
\begin{equation}
\bx^{m} = f(\bx^{m-1}) + \bT(\by^n-\bH(\bx^{m-1}))+ \hat{\bbeta}^m
\end{equation}
where we used time index $m$ for the state vector to emphasise that there are several model time steps between observation times.
$\bT$ is a relaxation matrix of our choice.
In this example, the deterministic part consists of the first two 
terms on the right-hand side of the equation, while the third term denotes the random part.
Let's assume the pdf of the random forcing is Gaussian with mean zero and covariance $\hat{\bQ}$. Then we can immediately 
write for the proposal density
\begin{equation}
q(\bx^m|\bx^{m-1},\by^n) = N\left(f(\bx^{m-1})+\bT(\by^n-\bH(\bx^{m-1})),\hat{\bQ}\right)
\end{equation}
since the pdf of $\bx^m$ is just a shift in the mean of the pdf of $\hat{\bbeta}^m$.
For the original model, we assume that the random part is Gaussian with zero mean and covariance $\bQ$,
so that
\begin{equation}
p(\bx^m|\bx^{m-1}) = N\left(f(\bx^{m-1}),\bQ\right)
\end{equation}

The change in the model equations is compensated for in particle filters by a change in the relative weight
of each particle, and the expression for this change in weight for this case is:
\begin{eqnarray}
w_i^m  & = &  w^{m-1}_i \frac{p(\bx_i^m|\bx_i^{m-1})}{q(\bx_i^m|\bx_i^{m-1},\by^n)} \nonumber \\
& \propto  & w_i^{m-1} \frac{\exp\left[ -J_p\right]}{\exp \left[ - J_q
             \right]} 
\end{eqnarray}
in which, for Gaussian model errors,
\begin{equation}
J_p =\frac{1}{2}\left(\bx_i^m-f(\bx_i^{m-1})\right)^T\bQ^{-1}\left(\bx_i^m-f(\bx_i^{m-1})\right)
\end{equation}
and 
\begin{eqnarray}
  J_q & = &\frac{1}{2}\left(\bx_i^m-f(\bx_i^{m-1})-\bT(\by^n-H(\bx^{m-1}))
            \right)^T \cdot \nonumber \\
& & \hat{\bQ}^{-1}\left(\bx_i^m-f(\bx_i^{m-1})-\bT(\by^n-\bH(\bx^{m-1}))\right) \nonumber \\
& = & \frac{1}{2}  (\hat{\bbeta}_i^m)^T  \hat{\bQ}^{-1}  \hat{\bbeta}_i^m
\end{eqnarray}
Note that the normalisation factors of the Gaussians do not have to be calculated explicitly
if we use that the sum of the weights has to be equal to one. The scheme is depicted by Algorithm \ref{alg:Relax}.

\begin{algorithm}[!ht]
	\caption{Relaxation Proposal Density \label{alg:Relax}}
	\begin{algorithmic}[0]
	    \For {$j = 1,...,N$}
		\State $\bd_j \gets  \by - H \left(\bx_j^f \right)$ 
		\State  $\bff_j \gets \bT\bd_j$  
                 \State $\bxi_j \sim N(0,\bQ)$  
		\State  $\bx_j^m \gets f\left(\bx_j^{m-1}\right) +\bff_j + \bxi_j$ 
		\State $\log w_j^m \gets \log w_j^{m-1} +
                  \frac{1}{2} \bxi_j\hat{\bQ}^{-1} \bxi_j $
                 \State $\log w_j^m \gets \log w_j^{m}-  \frac{1}{2}
                  (\bff_j+\bxi_j)^T \bQ^{-1}(\bff_j+ \bxi_j)$
            \EndFor
	\end{algorithmic}
\end{algorithm}

Simple as the scheme is, it does not solve the degeneracy problem. However, it can be used
as a simple scheme when several model time steps are used between observation times,
because the proposal is independent of the proposal at other time steps.
This scheme an easily be used in combination with other schemes that work at observation time,
to be discussed next.

\subsection{Weighted Ensemble Kalman Filter}
One could also use other existing data-assimilation methods in proposal densities, like
Ensemble Kalman filters or variational methods. In the Weighted Ensemble Kalman filter
\citep{Papadakis2010} the stochastic EnKF of \cite{Burgers98a} is used as follows. The Ensemble Kalman Filter 
update can be written as:
\begin{equation}
\bx_i^n = \bx_i^f + \bK(\by^n - \bH\bx_i^f - \bepsilon_i)
\end{equation}
in which $\bx_i^f = f(\bx_i^{n-1})+ \bbeta_i^n$, the matrix $K$ is the ensemble Kalman gain
and $\bepsilon_i \sim N(0,\bR)$, with $\bR$ the observational error covariance.
Using the expression for the forecast $\bx_i^f$ in the Kalman filter update equation we find:
\begin{equation}
\bx_i^n = f(\bx_i^{n-1}) + \bK\left(\by^n - \bH f(\bx_i^{n-1})\right)+ (\bI-\bK\bH)\bbeta_i^n - \bK \bepsilon_i
\end{equation}
which we can rewrite as the sum of a deterministic and a stochastic part as:
\begin{equation}
\bx^{n} = g(\bx^{n-1},\by^n) + \hat{\bbeta}_i^n
\end{equation}
identifying $g(\bx^{n-1}) = f(\bx_i^{n-1}) + \bK\left(\by^n - \bH f(\bx_i^{n-1})\right)$ and $\hat{\bbeta}_i^n = (\bI-\bK\bH)\bbeta_i^n - \bK \bepsilon_i$.
Therefore, we find for the proposal density:
\begin{equation}
q(\bx^n|\bx_i^{n-1},\by^n) = N\left(f(\bx^{n-1})+\bK(\by^n-\bH f(\bx^{n-1}),\hat{\bQ}\right)
\end{equation}
with 
\begin{equation}
\hat{\bQ} = (\bI-\bK\bH)\bQ(\bI-\bK\bH)^T + \bK\bR\bK^T .
\end{equation}
Strictly speaking, this is correct only if the Kalman gain is calculated using the ensemble covariance
of $f(\bx^{n-1})$, so without the model errors $\bbeta^n$, otherwise the proposal is not Gaussian.
We can calculate the weights of the particles in a similar way as in the previous example. 
Algorithm \ref{alg:WEKF} shows the algorithmic steps.

\begin{algorithm}[!ht]
	\caption{WEKF \label{alg:WEKF}}
	\begin{algorithmic}[0]
              \State $ \hat{\bQ} \gets (\bI-\bK\bH)\bQ(\bI-\bK\bH)^T + \bK\bR\bK^T $
	      \For {$i = 1,...,N$}
		    \State  $\hat{\bbeta}_i \sim N(0,\hat{\bQ})$  
                     \State $\bx_i^n \gets f(\bx_i^{n-1}) + \bK\left(\by^n - \bH f(\bx_i^{n-1})\right)+ \hat{\bbeta}_i^n $
	 	    \State  $w_i \gets  \frac{1}{2}\left(\bx_i^n-f(\bx_i^{n-1})\right)\bQ^{-1}\left(\bx_i^n-f(\bx_i^{n-1})\right)$
		    \State $w_i \gets  w_i +  \frac{1}{2}  \hat{\bbeta}_i  \hat{\bQ}^{-1}  \hat{\bbeta}_i $
		    \State $w_i \gets  w_i + \frac{1}{2} (\by - H(\bx_i^n))^T\bR^{-1}( \by - H(\bx_i^n)) $
		    \State $w_i \gets  \exp[-w_i] $
            \EndFor
            \State $\bw \gets  \bw / \bw^T \mathbf{1}$
            \State Resample
	\end{algorithmic}
\end{algorithm}

The behaviour of this filter has been studied extensively in \cite{Morzfeld2017}. In
high-dimensional systems this 
filter will be degenerate, consistent with the theory of \cite{Snyder2015}, and
as proven in the next section.
The only way to make this work is to include localisation, not only at the EnKF level, but also
at the level of the particle filter, see e.g. \cite{Morzfeld2017}. 

\subsection{Optimal proposal density}
\label{sec:optimal_proposal_density}

In the class of particle filters in which the proposal density of each particle is dependent on only
that particle, an optimal proposal density can be derived, as e.g. shown in \cite
{Doucet2001}. They defined optimality 
as the proposal density that gives a minimal variance of the weights, and
\cite{Snyder2015} provide an elegant proof of this optimality.
In this section we generalise this result and show that the
optimal proposal density is  
optimal even when each particle has its own proposal density which is
allowed to depend on all previous particles,
so a proposal of the form $q(\bx^n|i,\bx_{1:N}^{n-1},\by^n)$.

\cite{Snyder2015} concentrate on the case that one is interested in
an optimal  representation of $p(x^n,x^{n-1}|y^n)$ in a sequential
algorithm, so in a sequential smoother. 
To this end they introduce the random variable 
\begin{equation}
w^*(\bx^n,\bx^{n-1}) = \frac{p(\bx^n,\bx^{n-1}|\by^n)}{q(\bx^n,\bx^{n-1}|\by^n)}
\end{equation}
and determine that proposal density $q$ that minimises the variance in the weights $w^*$,
with the expectation taken over the density from which we draw the particles, so the proposal $q$.

Here we show that the optimal proposal density is also optimal for the strict filtering case,
so when we are interested in minimal variance of the weights at time $n$ only. 
Specifically, the question is: given the set of particles at $t=n-1$ drawn from $p(\bx_{n-1}|\by_{1:n-1})$,
which proposal density of the form $q(\bx^n|i,\bx_{1:N}^{n-1},\by^n)$ gives minimal variance of the weights at time $n$?

Using Bayes formula, we can write the expression 
for the weight of particle $i$ as function of the state at time $n$ as:
\begin{eqnarray}
\label{eq:opt-1}
w_i^n=w_i(\bx_i^n) & = & \frac{p(\by^n|\bx_i^n)}{N p(\by^n)}\frac{ p(\bx_i^n|\bx_i^{n-1}) }{  q(\bx_i^n|i,\bx_{1:N}^{n-1},\by^n)} \nonumber \\
& = & \frac{p(\by^n|\bx_i^{n-1})}{N p(\by^n)}\frac{ p(\bx_i^n|\bx_i^{n-1},\by^n) }{  q(\bx_i^n|i,\bx_{1:N}^{n-1},\by^n)}
\end{eqnarray}
where we assume, without loss of generality, an equally weighted ensemble at time $n-1$.
Note that the second equality follows from Bayes Theorem, as follows:
\begin{eqnarray}
p(\bx_i^n|\bx_i^{n-1},\by^n) & = &  \frac{p(\by^n|\bx_i^n,\bx_i^{n-1})}{p(\by^n|\bx_i^{n-1})} p(\bx_i^n|\bx_i^{n-1}) \nonumber \\
& = &  \frac{p(\by^n|\bx_i^n)}{p(\by^n|\bx_i^{n-1})} p(\bx_i^n|\bx_i^{n-1})
\end{eqnarray}

Consider the pair of random variables $(I,\bX^n)$ where $Prob(I=i) =
\frac{1}{N}$ and, conditionally on $I=i$, $\bX^n \sim q(\bx^n|i,\bx_{1:N}^{n-1},\by^n)$.
Furthermore, define the associated random variable
\begin{equation}
W=w_I(\bX^n) = \frac{p(\by^n|\bx_I^{n-1})}{Np(\by^n)}\frac{ p(\bX^n|\bx_I^{n-1},\by^n) }
{q(\bX^n|I,\bx_{1:N}^{n-1},\by^n)}
\end{equation}
where 
\begin{equation}
p(\by^n) = \frac{1}{N} \sum_{j=1}^N p(\by^n|\bx_j^{n-1})
\end{equation}
In order to find the proposal $q$ that minimizes the variance of $W$,
we use the well-known law of total variance (derived in the appendix for completeness):
\begin{equation}
var_W(W) = var_I(E_{\bX^n|I}(W)) + E_I(var_{\bX^n|I}(W)).
\end{equation}
First, we see that, under the proposal $q$:
\begin{equation}
E_{\bX^n|I}(W) = \frac{p(\by^n|\bx_I^{n-1})}{Np(\by^n)} \int p(\bx^n|\bx_I^{n-1},\by^n)
d\bx^n = \frac{p(\by^n|\bx_I^{n-1})}{Np(\by^n)}
\end{equation}
is independent of $q$. Moreover, $E_W(W)=E_I(E_{\bX^n|I}(W))=1/N$ and thus
the first term in $var_W(W)$ is
\begin{equation}
\frac{1}{N} \sum_i \frac{p(\by^n|\bx_i^{n-1})^2}{N^2p(\by^n)^2} -
\frac{1}{N^2}
=\frac{1}{N} \sum_i \left(\frac{p(\by^n|\bx_i^{n-1})}{Np(\by^n)} -
  \frac{1}{N} \right) ^2 \geq 0.
\end{equation}
For the second term
we use that $var_{\bX^n|I}(W)\geq 0$ with equality 
if and only if $W$ is almost surely constant in $\bX^n$,
that is if and only if
\begin{equation}
\frac{ p(\bx^n|\bx_i^{n-1},\by^n) }{q(\bx^n|i,\bx_{1:N}^{n-1},\by^n)} = cst(i,\bx_{1:N}^{n-1},\by^n).
\end{equation}
in which $cst(..)$ is this constant which can depend on other
    variables than $\bx^n$.
Because both $p$ and $q$ are densities (in $\bx^n$), $cst=1$. Combining
these results, we have a lower bound for $var(W)$ that is determined by the variance
of $p(\by^n|\bx_i^{n-1})$ over $i$, with equality if and only if
\begin{equation}
q(\bx^n|i,\bx_{1:N}^{n-1},\by^n)=p(\bx^n|\bx_i^{n-1},\by^n)
\end{equation}
Note that this is a new result as previous proofs only considered proposal densities of the form $q(\bx^n|\bx_i^{n-1},\by^n)$,
and we extended it to more general proposal densities of the form  $q(\bx^n|i,\bx_{1:N}^{n-1},\by^n)$.
  
This remarkable result shows that firstly the optimal proposal density, so 
$p(\bx^n|\bx_i^{n-1},\by^n)$, does indeed lead to the lowest variance in the weights 
for the class of particle filters in which the transition density is of the form $q(\bx^n|i,\bx_{1:N}^{n-1},\by^n)$.
Secondly, it shows that we can predict the variance in the weights without doing the 
actual experiment, for any number of particles, provided we can compute
$p(\by^n|\bx^{n-1}_i)$, and thirdly the weights are {\em independent of the position
of the particles $\bx^n$}. Unfortunately, this variance is zero only 
when the observations are not dependent on the state at time $n-1$, which
is never the case in the geosciences.

A simple case where we can compute both the optimal proposal density and
the weights $p(\by^n|\bx^{n-1}_i)$ is when $p(\bx^n|\bx^{n-1}_i)$ is given
by \eqref{eq:model-Gauss} and the observation operator $H =\bH$ is linear. By
the same argument that 
is used to derive the Kalman filter update, we find
\begin{equation}
p(\bx^n|\bx^{n-1}_i,\by^n) = \nonumber
\end{equation}
\begin{equation}
= N\left(f(\bx^{n-1}_i) + 
\bT(\by^n -\bH f(\bx^{n-1}_i)),(\bI-\bT \bH^T)\bQ\right),
\end{equation}
where $\bT = \bQ\bH^T(\bH\bQ\bH^T+\bR)^{-1}$ is the Kalman-like gain with the background covariance $\bQ$, and
the weights are proportional to:
\begin{equation}
p(\by^n|\bx^{n-1}_i) = N(\bH f(\bx^{n-1}_i), \bH\bQ \bH^T + \bR)
\end{equation}
This shows two things: First, in this special case, the simple relaxation
scheme of Section 
2.1 is equal to the optimal proposal when the relaxation matrix $\bT$ is chosen as 
above. Second, comparing the weights of the optimal proposal
with the weights of the standard filter, they both depend on the squared distance
$||\by^n - \bH f(\bx^{n-1}_i)||^2$, and $||\by^n - \bH \bx^{n}_i||^2$, respectively, but in the standard particle filter the
distance is defined w.r. to $\bR$ and in the optimal proposal the distance 
it is defined is w.r.
to $\bH \bQ \bH^T + \bR$. Hence the weights with the optimal proposal are more similar,
but the improvement is substantial only if $\bQ$ is large, and the analysis
of  weight collapse by \cite{Snyder2008}  still applies. 

One can extend the optimal proposal density idea to more than one time step.
\cite{Snyder2015} show that the optimal proposal is the proposal of this form with minimal variance in the 
weights in this case too, which can also easily be seen by applying the above to
$$W=w_i(\bx^n) = \frac{p(\by^n|\bx_i^{m-1})}{Np(\by^n)}\frac{ p(\bx^n|\bx_i^{m-1},\by^n) }
{q(\bx^n|\bx_i^{m-1},\by^n)}$$
for $m<n$.

Looking back at the filters described in the previous sections we find the following.
The relaxation scheme uses a simple proposal density that is of the form $q(\bx^n|\bx_i^{n-1},\by^n)$,
so the theory holds, and that proposal will lead to degenerate results.
This is indeed the finding of \cite{VanLeeuwen2010}.
The Weighted Ensemble Kalman Filter has a proposal that depends on all particles at time $n-1$ through
 the Kalman gain $\bK$, so the proposal is of the form $q(\bx^n|i,\bx_{1:N}^{n-1},\by^n)$.
Hence also this filter will perform worse than the optimal proposal and hence will be degenerate for high-dimensional 
systems. This was first explored in detail by \cite{Morzfeld2017}.

\subsection{Implicit Particle filter }

The Implicit Particle Filter is an indirect way to draw from the optimal proposal, even over several time steps.
Often the assumption is made that the model errors of both original model and proposal density
are Gaussian, and the observation operator $\bH$ is linear. In this case, a draw from the optimal
proposal is a draw from a multivariate Gaussian, and we know how to do that.

\begin{figure}[h]
	\centering\includegraphics[width=1\linewidth]{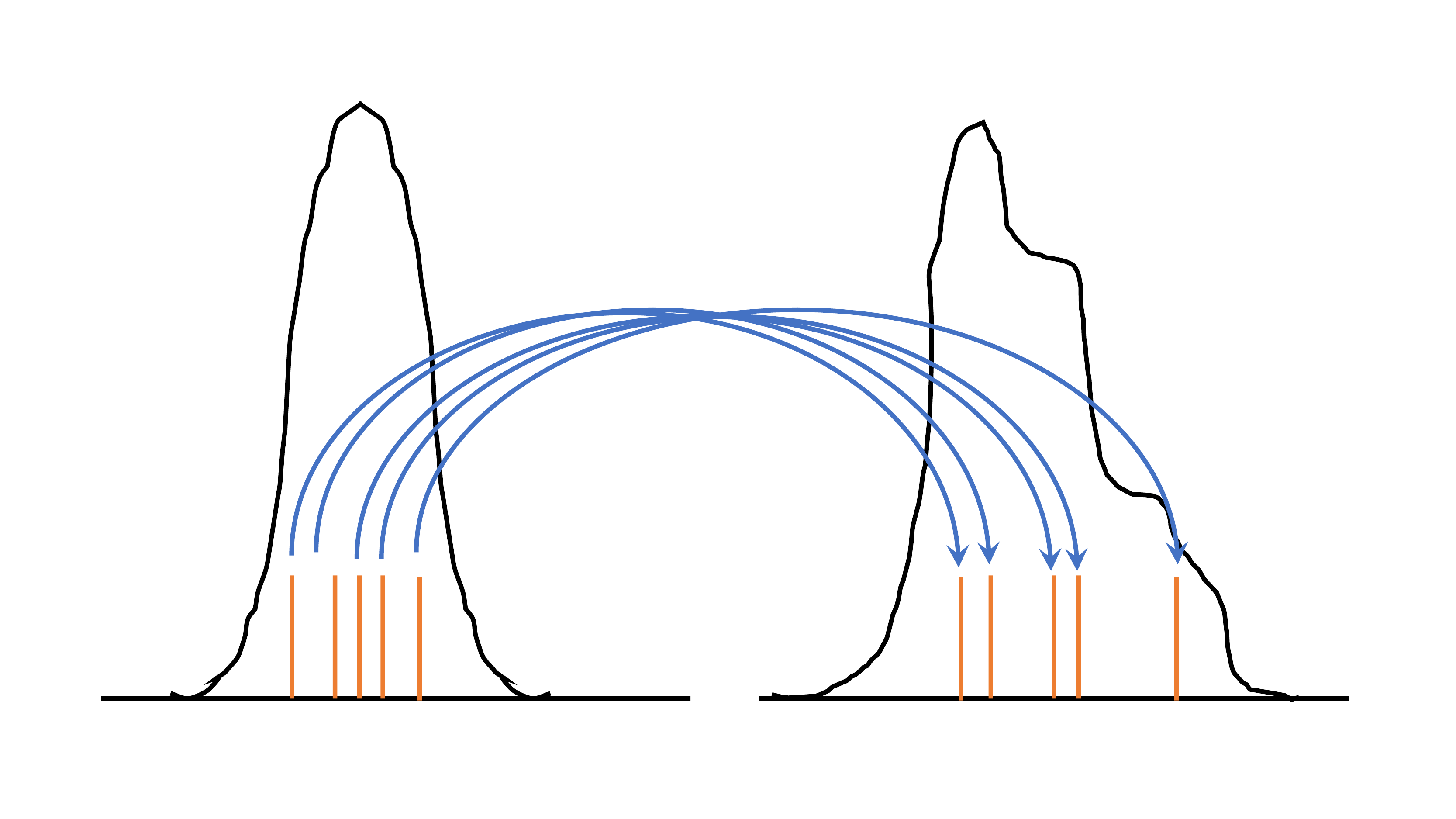}   
	\caption{The Implicit Particle Filter. Samples (red bars in left pdf) are drawn from the standard multivariate Gaussian
	and transformed via equation (\ref{eq:implicit}) to weighted samples from the posterior (red bars in right pdf). } 
	\label{fig:IPF}
\end{figure}

However, when $\bH$ is nonlinear, or when the proposal is used over several model time steps 
the density to draw from is not Gaussian anymore. \cite{Chorin2010} realised that one could
still draw from a Gaussian and then apply a transformation to that draw to find samples from the 
optimal proposal density. 
The method is explained here for one time step, but the extension to multiple time steps is straightforward.
Figure \ref{fig:IPF} illustrates the basic idea.

As mentioned in Sec.\ \ref{sec:proposal_density_pfs} on the proposal density the posterior pdf can be written as:
\begin{equation}
p(\bx^n|\by^n) = \sum_{i=1}^N w^{n-1}_i \frac{p(\by^n|\bx^n)}{p(\by^n)}\frac{p(\bx^n|\bx_i^{n-1})}{q(\bx^n|\bx_i^{n-1},\by^n)}q(\bx^n|\bx_i^{n-1},\by^n)
\end{equation}
The scheme draws from a Gaussian proposal $q(\bxi) = N(0,\bI)$, and we can write the transformation as 
$q(\bx^n|\bx_i^{n-1},\by^n) = q(\bxi) \bJ_i^{-1}$ in which $\bJ_i$ is the Jacobian of the transformation from $\bx^n$ to $\bxi$.
That transformation is found implicitly, hence the name of the filter, by defining
\begin{equation}
F_i(\bx^n) = - \log\left[ p(\by^n|\bx^n)p(\bx^n|\bx_i^{n-1}) \right]
\end{equation}
 and, after drawing $\bxi_i$ for each particle, solving for $\bx^n$ in
\begin{equation}
F_i(\bx^n) =\frac{1}{2} \bxi_i^T \bxi_i + \phi_i
\label{eq:implicit}
\end{equation}
for each particle, in which $\phi_i = \min_{\bx^n} F_i(\bx^n) \propto p(\by^n|\bx_i^{n-1})$.
The weights of the particles become:
\begin{eqnarray}
w_i^n & = &  w^{n-1}_i \frac{p(\by^n|\bx_i^n)}{p(\by^n)}\frac{p(\bx_i^n|\bx_i^{n-1})}{q(\bx_i^n|\bx_i^{n-1}\by^n)}  \nonumber \\
& = & w^{n-1}_i \frac{\exp\left[-F_i(\bx_i^n)\right]}{\exp\left[ -\frac{1}{2}\bxi_i^T \bxi_i \right]} \bJ_i \nonumber \\
& = & w^{n-1}_i \frac{\exp\left[-F_i(\bx_i^n)\right]}{\exp\left[ -F_i(\bx^n) + \phi_i \right]} \bJ_i \nonumber \\
& = & w^{n-1}_i \exp\left[ - \phi_i \right] \bJ_i 
\end{eqnarray}
Interestingly, while the optimal proposal density shows that the weights are only dependent on
the position of the particles at the previous time, so on $\bx_i^{n-1}$ via $\phi_i$, the
implicit map makes the weights also dependent on the positions at the current time $n$, so on $\bx_i^n$
via the Jacobian of the transformation between $\bxi$ and $\bx$. 
Only when the Jacobian is a constant, so when $F_i$ is quadratic in $\bx_i$, this
dependence disappears.

Solving (\ref{eq:implicit}) is not straightforward in general.
\cite{Morzfeld2012} suggest a random map of the form
\begin{equation}
\bx_i^n = \bx_i^a + \lambda_i(\bxi_i) \bP^{1/2} \bxi_i
\end{equation}
in which $\bP$ is a chosen covariance matrix, ideally the covariance of the posterior pdf, 
$\bx_i^a = \arg \min F_i(\bx^n)$ and $\lambda_i$ is a scalar. This transforms the problem into 
solving a highly nonlinear scalar equation for $\lambda_i$, which is a much simpler problem than 
finding $\bx_i^n$ directly. This map can be shown to be a bijection when $F_i(\bx_i^n)$ has only closed
contours in the high-probability regions; otherwise one would have to first choose a closed 
contour area and then perform the map. 
In general, when the optimal proposal (over several time steps if needed) is multimodal, the transformation from 
the state variable to a Gaussian is not monotonic, and the Implicit Particle Filter needs to be adapted, e.g. by
using a separate Gaussian for each mode.
The algorithm is given in Algorithm \ref{alg:IPF}.

\begin{algorithm}[!ht]
	\caption{Implicit Particle Filter \label{alg:IPF}}
	\begin{algorithmic}[0]
	    \For {$i = 1,...,N$}
		\State $\bxi_i \sim N(0,\bI)$ 
		\State  $\phi_i \gets \min_{\bx^n} \{- \log\left[ p(\by^n|\bx^n)p(\bx^n|\bx_i^{n-1}) \right] \}$  
                 \State Solve $- \log\left[ p(\by^n|\bx^n)p(\bx^n|\bx_i^{n-1}) \right] = \frac{1}{2} \bxi_i^T \bxi_i + \phi_i$ for $\bx^n$
                 \State $ \bJ_i = \left | \frac{\partial \bx^n}{\partial \bxi_i} \right | $
		\State $w_i \gets \exp\left[ - \phi_i \right] \bJ_i$ 
            \EndFor
            \State $\bw \gets \bw / \bw^T \mathbf{1} $
            \State Resample
	\end{algorithmic}
\end{algorithm}

Of further interest is that $\bx_i^a$ is the same as the solution to a 4DVar problem well known in meteorology.
But it is a special 4DVar 
as the initial position of each particle is fixed and it has to be a weak-constraint 4DVar.
The latter condition is needed as a strong-constraint 4Dvar would have no possibility to move a 
particle in state space as its initial condition is fixed. 

However, also this filter will suffer from weight collapse in high-dimensional applications
as it is still a sampling scheme for the optimal proposal density. 
The following sections will discuss ways to improve on the optimal proposal.

\subsection{Equal weights by resampling at time $n-1$}
As noted already in equation \eqref{eq:opt-1}, we can write equation
\eqref{eq:mix-posterior} as
\begin{eqnarray}
p(\bx^n|\by^n) &=& \sum_{i=1}^N \frac{w^{n-1}_i
  p(\by^n|\bx^{n-1}_i)}{p(\by^n)} p(\bx^n|\bx^{n-1}_i,\by^n) \nonumber \\ 
  & = & \sum_{i=1}^N \alpha_i p(\bx^n|\bx_i^{n-1},\by^n) \label{eq:mixt}
\end{eqnarray}
where
\begin{equation}
\alpha_i = \frac{ w^{n-1}_i  p(\by^n|\bx_i^{n-1})}{p(\by^n)}
\end{equation}
This says that, assuming the pdf at the previous time can be approximated by
a set of $N$ particles, the analysis distribution is a mixture of the
optimal proposal pdf's $p(\bx^n|\bx_i^{n-1},\by^n)$ with mixture weights
$\alpha_i$. 

If we can compute the optimal proposal density and the weights $\alpha_i$
in closed form, 
we can also draw samples {\em directly from this mixture density}. For this,
we first draw an index $I$  from the discrete distribution
with weights $\alpha_i$, $Prob(I=j) = \alpha_j$, followed by a draw from
the corresponding  
pdf $p(\bx^n|\bx_I^{n-1},\by^n)$. Doing this $N$ times will lead to $N$
different particles with equal weights because each of them 
is an independent draw directly from the posterior. If the index $I$ 
is equal to a value $j$ more than once, the particle $\bx^{n-1}_j$ 
is propagated from time $n-1$ to time $n$ with independent random forcing
for each of these draws. This simple scheme provides better
samples than the optimal proposal density because all particles are different
at time $n$ by construction.

However, this does not solve the problem of weight collapse because drawing
the index $I$ is nothing else than resampling the particles at time $n-1$
with weights proportional to $w^{n-1}_i  p(\by^n|\bx_i^{n-1})$. If
$w^{n-1}_i=\frac{1}{N}$, the variance of these weights is exactly equal to
the lower bound that we found in Section 2.3. The main difference is that
the collapse now happens at time $n-1$. The only advantage is that 
all particles will be different at time $n$.

If we cannot compute the optimal proposal density and the weights
$\alpha_i$ in closed form, we can still use the importance
sampling idea to draw from the mixture $p(\bx^n|\by^n)$ 
by drawing pairs $(I,\bX^n)$ consisting of an index $I$ and a
state $\bX^n$ at time $n$. We choose a proposal distribution 
$\beta_i = \beta_i(\by^n)$ for the index and proposal distributions
$q(\bx^n|\bx^{n-1}_i,\by^n)$ for the state. Then we draw the index $I_i$
with $Prob(I_i=j)=\beta_j(\by^n)$ and conditionally on $I_i=j$ we
draw $\bx^n_i$ from $q(\bx^n|\bx_j^{n-1},\by^n)$. Finally, we compute 
weights $w^n_i$ by
$$w^n_i \propto \frac{w^{n-1}_{j} p(\bx^n_i|\bx^{n-1}_{j})p(\by|\bx^n_i)}
  {\beta_{j}(\by^n) q(\bx^n|\bx_{j}^{n-1},\by^n)}  \textrm{ if } I_i=j$$
The particles $\bx^n_i$ with weights $w^n_i$ provide the desired
approximation of $p(\bx^n|\by^n)$ whereas the indices $I_i$ can be
discarded after the weights have been computed. We could produce
an evenly-weighted approximation by a further resampling step, or
take the weights $w^n_i$ into account during the next iteration.

In this approach we can obtain equal weights $w^n_i$ by choosing
$$q(\bx^n|\bx_{j}^{n-1},\by^n)= p(\bx^n|\bx_{j}^{n-1},\by^n)$$
and 
$$\beta_i(\by^n) \propto w^{n-1}_i p(\by^n|\bx^{n-1}_i).$$
With this choice, we draw directly from the mixture (\ref{eq:mixt}).
As mentioned before, although the weights $w^n_i$ are then equal to 
$\frac{1}{N}$, the algorithm contains a hidden weighting and resampling
step of particles at time $n-1$. It thus remains susceptible to weight
collapse in high dimensions.  

This approach of using importance sampling for the joint distribution of
$(I,\bX^n)$ is due to \cite{Pitt1999} who called it ``Auxiliary Particle
Filter'' (the index $I$ is an auxiliary variable that is discarded at the
end). They discuss, in addition, approximations of the
optimal proposal density and the optimal weights $\alpha_i$. One of their
suggestions is to use for the index $I$ the proposal with weights
$$\beta_i \propto w^{n-1}_i p(\by^n| \bmu^n_i)$$
where $\bmu^n_i$ is a likely value of the distribution
$p(\bx^n|\bx^{n-1}_i)$, e.g. the mean or median or simply a draw from it.
Typically, $\bmu^n_i$ is found by a probing step where particles at time
$n$ are propagated by a simplified model, e.g. by omitting
stochastic terms or with simplified subgrid-scale parameterisations or 
thermodynamics. If $I_i=j$ and the state $\bx^n_i$ at time $n$ is proposed from
$p(\bx^n| \bx^{n-1}_j)$, the weights become
$$w^n_i \propto \frac{p(\by|\bx^n_i)} {p(\by^n| \bmu^n_j)}$$
They will vary less provided $\bx^n_i$ is close to $\bmu^n_j$, i.e.
provided the simplified model is a good approximation to the full model
and the stochastic part of the full model is small.

\subsection{The Equivalent-Weights Particle Filter (EWPF)}
The EWPF \citep{VanLeeuwen2010,Ades13} uses the idea 
to obtain a more evenly weighted set of particles by not sampling
from the exact posterior, but allowing for a small error. It
starts with determining the weight of each particle at the mode of
$p(\bx^n|\bx_i^{n-1},\by^n)$
for each particle $i$, $w_i^{max} \propto p(\by^n|\bx_i^{n-1})$.
Note that these weights are equal to the weights obtained in the optimal proposal density.
In the optimal proposal density case the weights do not depend on the position $\bx^n$ of the particle,
but note that the proposal used here will be different.

The particles are not moved to these modes, but the weights are used to define a target weight.
This target weight $w_{target}$ is chosen such that a certain fraction $\rho$ of particles can reach that weight.
To this end we sort the weights in magnitude from high to low in an array $w^*_i,\; i = \{1,2,...,N]\}$ and set  $w_{target} = w^*_{N*\rho}$.
For instance, with 100 particles and a fraction of $\rho = 0.8 $ we would find $w_{target} = w^*_{80}$.

The next step is to find a position in state space for each particle that can reach this weight
such that its weight is exactly equal to the target weight. This means we solve for $\bx^n$ in
\begin{equation}
w_i(\bx^n) = w_{target}
\label{eq:target}
\end{equation}
for each particle $i$ that can reach this weight. There are many solutions
of this equation, but we choose the one which is on the line through
$\bx^a_i$ and $f(\bx^{n-1}_i)$ and is closest to $f(\bx^{n-1}_i)$.
Denote this position as $\bx_i^*$. Note that this is purely deterministic move, so a stochastic part still has to be added.
The final position of these particles is then determined by adding a very small random perturbation $\bxi$
from a chosen density, so
\begin{equation}
\bx_i^n = \bx_i^* + \bxi_i^n
\end{equation}
This stochastic move  ensures that the proposal has full support and is not
a delta function centred at $\bx_i^*$. The density of $\bxi_i$ should on
the one hand have most of its mass concentrated around 0 in order 
not to change the weights of the particles too much, and on the other hand
it should be relatively constant since we divide by the value of the
proposal density. Both requirements cannot be fulfilled exactly, but
we can take some error in the sampling into account and choose 
a narrow uniform distribution. The scheme is depicted in Algorithm
\ref{alg:EWPF} for the special case that of Gaussian model errors and a linear observation operator.
If these conditions do not hold, one will typically need iterations 
to solve for $a_i$ and $b_i$.

\begin{algorithm}[!ht]
	\caption{EWPF \label{alg:EWPF}}
	\begin{algorithmic}[0]
		\State $\epsilon \gets 0.0001/N$
		\State $\gamma_U \gets 10^{-6}$ 		
		\State $\gamma_N \gets \frac{2^{N_x/2}\epsilon \gamma^{N_x}_U}{\pi^{N_x/2} (1-\epsilon)}$ 	
		\State $N_{k} \gets N  \rho$ 
		\For {$j = 1,...,N$} 
			\State $\bd_j \gets  \by - \bH \left(f(\bx_j^{m-1}) \right)$ 
			\State $c_j \gets -\log\bw^{m-1} + 0.5\bd_j^T \left(\bH\bQ\bH^T+\bR \right)^{-1} \bd_j$
		\EndFor
		\State $\left( \hat{\bc},  \bidx \right)  \gets sort(\bc) $ 
		\State $C_{max} \gets \hat{\bc}(N_{k}) $ 
		\For{$j = 1,...,N_{k}$} 
			\State $i \gets \bidx(j)$ 	
	         	\State $a_i \gets \frac{1}{2} \bd_i^T\bR^{-1} \bH \bQ \bH^T \left(\bH\bQ\bH^T+\bR \right)^{-1} \bd_j$     
	         	\State $b_i \gets \frac{1}{2} \bd_i^T\bR^{-1}  \bd_i - C_{max} -\log \bw^{m-1}$   
	         	\State  $\alpha_i \gets 1+ \sqrt{1-b_i/a_i}$        	
			\State $\bbeta_i \sim (1-\epsilon)\bQ^{1/2} \bU \left(-\gamma_U \bI,+\gamma_U \bI  \right) + \epsilon N \left(\gamma^2_N \bQ \right)$   
			\State $\bx_j^a \gets f \left(\bx_i^{m-1}\right) + \alpha_i \bQ \bH^T \left(\bH\bQ\bH^T+\bR \right)^{-1} \bd_j +\bbeta_i $ 
			\If{$\bbeta_i$ was from uniform distribution}
				\State $\tilde{c}_j \gets -\log \bw_i^{m-1} + (\alpha_i^2 - 2 \alpha_i) a_i + \frac{1}{2} \bd_i^T\bR^{-1}  \bd_i $  
			\Else
				\State $v_1 \gets -\log \bw_i^{m-1} + (\alpha_i^2 - 2 \alpha_i) a_i $ 
				\State $v_2 \gets v_1 + \frac{1}{2} \bd_i^T\bR^{-1}  \bd_i \left(2^{-N_x/2}\right) \left(\pi^{N_x/2}\right) $ 
				\State $v_3 \gets v_2  \gamma_N \gamma_U^{-N_x} (\frac{1-\epsilon}{\epsilon})  $ 
				\State $\tilde{c}_j \gets v_3  \exp \left( 0.5 \bbeta_i^2  \right) $  
			\EndIf
		\EndFor
		\State $\bw = \exp(-\tilde{\bc})$
	         \State $ \bw \gets \bw / \bw^T \mathbf{1} $
		\State Resample to have full ensemble, $\bX^a$, of $N$ particles from $N_k$ particles  $\bx^a$.  
	\end{algorithmic}
\end{algorithm}

It is common knowledge, see e.g. \cite{Doucet2001}, that the proposal should be wider or at least as wide as the
target, while the width of the stochastic part of the proposal is chosen very small here.
The reason that we can do this is that the position of the centres of these proposal densities
are typically further away from the observations than e.g. in the optimal proposal
because the target weight forces particles away from their optimal positions, so away from the observations.
This means that the deterministic moves of the particles ensure a large spread in the full proposal.

A formal way to avoid such an error has been described by 
\cite{Ades2015a}. They choose the proposal to be a mixture of a uniform density and a Gaussian which is also used in Alg.\ \ref{alg:EWPF}. 
Both have small variance, and the mixture coefficient of the uniform density is chosen to be much larger 
than that of the Gaussian. This means that drawing from the Gaussian and also drawing from its tails becomes
highly unlikely. In practice, since we always work with small ensemble sizes the chance of filter degeneracy
by drawing from the Gaussian, and then drawing from the tail of the Gaussian is indeed highly unlikely.

Finally, the full weights for the new particles are calculated and the whole ensemble is resampled, including
those particles that were unable to reach the target weight. Because of the target-weight construction 
the weights of the particles are very similar, and filter degeneracy is avoided.
This filter has been used in a reduced-gravity ocean model by \cite{Ades2015a},
and in the same system studied for the gravity-wave production by the scheme in \cite{Ades2015b}.
It has also been applied in a climate model by \cite{Browne2015b}.

To analyse the scheme further, we can again look at the variance of the weights.
For this it is important to note that this scheme does not see the weight of a particle as a 
function of the state $\bX$ and particle index $I$, but rather the state as function of the weight $W$
and index $I$, so $\bX(W,I)$. 
Specifically, $W|I$ has values in two ranges. For the particles with $I=i$
that can reach the target weight we find $w|I = w_{target} + \epsilon_i$ in which $\epsilon_i$ is a small perturbation 
from the target weight due to the small stochastic move discussed above.
For those particles that cannot reach the target weight their weights are very close to zero.
So we find:
\begin{equation}
E_I[W] \approx \rho (w_{target}+\bar{\epsilon})+(1-\rho)0 = \rho (w_{target}+\bar{\epsilon})
\end{equation}
in which $\bar{\epsilon}=E_I[\epsilon]$.
If $H$ is linear and the errors in the observations and the model equations are Gaussian we find
$\bar{\epsilon}=0$, but if any of these three conditions does not hold this is not necessarily so.
However, we do know that by construction $|\bar{\epsilon}|<<1$.
Since the sum of the weights should be equal to 1, we find that $w_{target}\approx1/(N\rho)$,
and hence $E_I[W]=1/N$, as expected.
Furthermore
\begin{eqnarray}
var_I(W) & = &  \rho \sum_{i=1}^{\rho N}(w_{target}+\epsilon_i)^2  - (\rho w_{target})^2  \nonumber \\
 & \approx  & \frac{1}{N^2}\frac{1-\rho}{\rho}
 \end{eqnarray}
This expression shows that the variance in the weights ranges between
$0$ for $\rho=1$, so when all particles are kept, to $(N-1)/N^2 \approx 1/N$ for $\rho=1/N$,
so when one particle is kept. 
We can compare this with the optimal proposal when the number of independent observations is large.
In that case one particle will have a weight very close to one, and the rest will have weights very
close to zero. The variance in the weights is then $(N-1)/N^2 \approx 1/N$, indeed equal to
the $\rho=1/N$ case in the EWPF scheme, as expected. The EWPF can, however, reduce that
variance, even to zero, depending on the choice of the tuning parameter $\rho$. 

When this tuning parameter is chosen close to one, the target weight will be low, and hence
particles will be moved further away from the mode of the optimal proposal density.
In practise this means that the particles are pushed further away from each other, leading to
a wider posterior pdf. A small value for the fraction will have the opposite effect.  
Since we do not know a-priori what the width of the posterior should be, this is a clear drawback of this method.
We will come back to this later.

\subsection{The Implicit Equal-Weights Particle Filter}

In the Implicit Equal Weights Particle Filter (IEWPF) we set the target weight equal to the minimum of the optimal
proposal weights for all particles. Then, the position of each particle is set to the mode of the optimal proposal 
density plus a scaled random perturbation. The scale factor is chosen such that the weight of each particle is equal to the target weight.
Note that in the standard setting no resampling is needed, but see \cite{Zhu16} for other possibilities.

The implicit part of the scheme follows from drawing samples implicitly from a standard Gaussian distributed proposal 
density $q(\bxi)$ instead of the original $q(\bx^n|\bx^{n-1},\by^n)$, following the same procedure as in the Implicit Particle Filter.
We define a relation
\begin{equation} 
\bx_i^n = \bx_i^a + {\alpha}_i^{1/2} \bP^{1/2} \bxi_i^n
\end{equation}
where $\bx_i^a$ is the mode of $p(\bx^n|\bx_i^{n-1},\by^n)$, $\bP$ is a measure of the width of that pdf, $\bxi_i^n \in \Re^{N_x}$ is a standard Gaussian-distributed random
vector, and ${{\alpha}_i}$ is a scalar. 

The IEWPF scheme is different from the Implicit Particle Filter in that it chooses the ${{\alpha}_i}$ such that all particles get the same weight ${w_{target}}$, so the scalar ${{\alpha}_i}$ is determined for each particle from:
\begin{equation}
      w_i(\alpha_i)=\frac{   p(\bx_i^n|\bx_i^{n-1},\by^n)p(\by^n|\bx_i^{n-1})}{Np(\by^n)q(\bx^n|i,\bx_{1:N}^{n-1},\by^n)}   = w_{target}
\end{equation}
This target weight is equal to the lowest weight over all particles in an optimal proposal.
This ensures that the filter is not degenerate in systems with arbitrary dimensions and an arbitrary number of independent observations.
The resulting equation for each $\alpha_i$ is nonlinear and complex because it will contain the Jacobian of the transformation
from $\bxi^n$ to $\bx^n$, similar to the Implicit Particle Filter. The Jacobian will contain
the derivative of $\alpha_i$ to $\bxi_i$, which is the main source of the complexity in this scheme.
Algorithm \ref{alg:IEWPF} depicts the scheme for the case of a linear observation operator.
A nonlinear observation operator will lead to more complicated equations for the $\alpha$'s.

\begin{algorithm}[!ht]
	\caption{IEWPF \label{alg:IEWPF}}
	\begin{algorithmic}[0]
		\For {$j = 1,...,N$}  
			\State $\bd_j \gets  \by - \bH \left(f(\bx_j^{m-1}) \right)$ 
			\State $c_j \gets -log \bw^{m-1} + 0.5\bd_j^T \left(\bH\bQ\bH^T+\bR \right)^{-1} \bd_j$
		\EndFor
		\State $c_{target} \gets \min \left(\bc \right)$ 
	         \State $\bP \gets (\bQ^{-1}+\bH^T\bR^{-1}\bH)^{-1}$ 
	         \State $\bxi_i \sim N(0,\bP)$ 
		\For{$j = 1,...,N$} 
	         	\State  $\bx_j^a \gets f\left(\bx_j^{m-1}\right) + \bQ\bH^T \left(\bH\bQ \bH^T+\bR \right)^{-1} \bd_j$ 
			\State $\gamma_j \gets \bxi_j^T \bxi_j $   
			\State $a_j \gets \bd_j^T \left(\bH\bQ \bH^T+\bR \right)^{-1} \bd_j +\log \bw^{m-1}+ c_{target}$
			\State Solve $(\alpha_j-1) {\gamma}_j - N_x \log \alpha_j + a_j=0$ for $\alpha_j$ 
			\State $\bx_j^n \gets \bx_j^a + \alpha_j \bxi_j$  
		\EndFor
	\end{algorithmic}
\end{algorithm}

The scheme is similar to the optimal proposal density using the Implicit Particle Filter by first determining the mode
of the proposal and then adding a random vector. The difference is that in the IEWPF the size of the vector is
determined such that the each particle reaches the target weight. It turns out that this construction excludes part
of state space for all but one particle. For each particle the excluded part is different, so the ensemble samples the 
whole space, but the individual particles do not.
Details of the method can be found in \cite{Zhu16}.

\begin{figure}[h]
	\centering\includegraphics[width=1\linewidth]{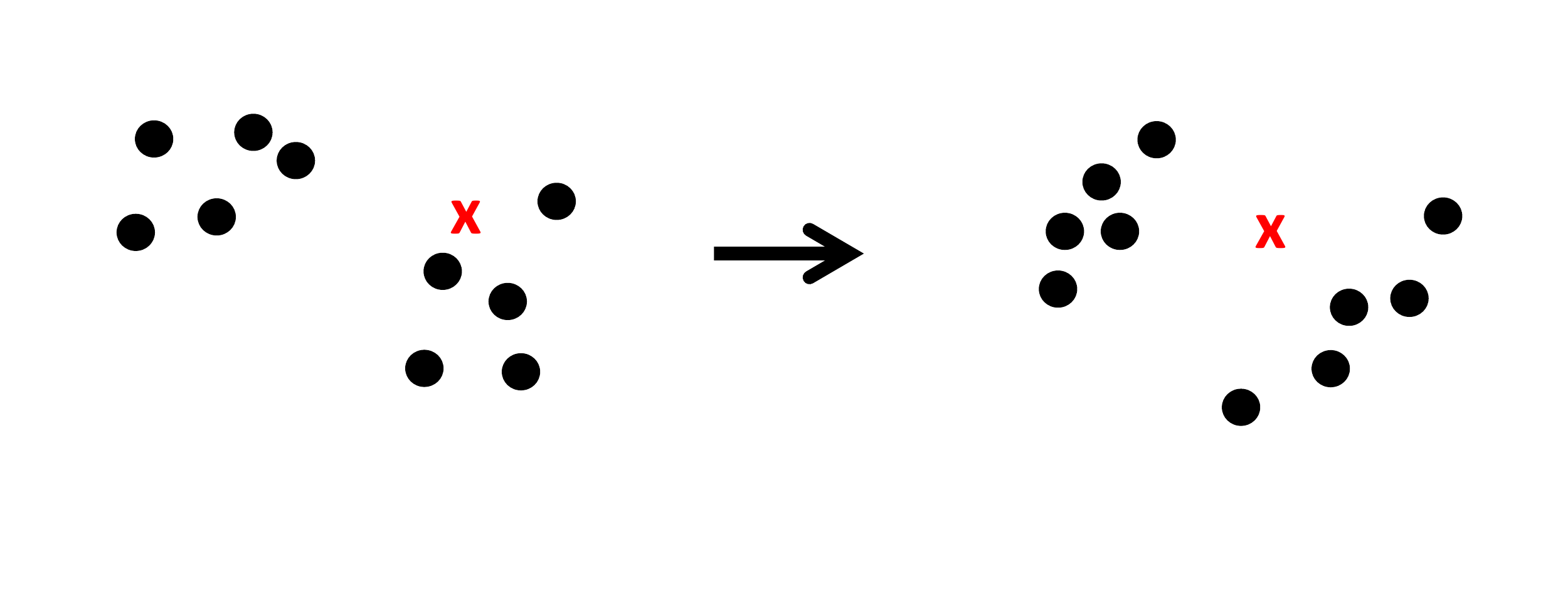}   
	\caption{The implicit equal-weights particle filter. Left: the prior particles at time $n-1$(dots), with one observation,
	denoted with the red cross. Right: the posterior particles. Note that the weights are equal,
	but some particles have moved away from the observations to ensure equal weights.} 
	\label{fig:Equal-Weight-PF}
\end{figure}

Analysing the scheme in more detail, the proposal density used in this scheme is of one dimension lower than that
of the state itself. The {\em direction} of the random vector in state space is determined by the proposal density, but 
the {\em size} of the random vector is then determined deterministically, dependent on that direction. So the proposal
density misses one degree of freedom for all but one particle, the particle with the lowest weight that has $\alpha_i=1$.
Although missing one degree of freedom in a very high dimensional system might seem acceptable it does lead
to a bias. 
Figure \ref{fig:Equal-Weight-PF} shows how the implicit equal-weights particle filter works.

\subsection{Discussion}

We first note that the optimal proposal is only optimal in a very limited sense, as has been known a long time
with the invention of the auxiliary particle filter.
We have seen that it is not difficult to generate particle filters
that even have zero variance in the weights.
In the optimal proposal setting one forces $Prob(I=i)=1/N$, while the simple choice $Prob(I=i)\propto
p(y^n|x_i^{n-1})$ leads to an equal-weight particle filter. 
Furthermore, schemes have been introduced that consider the state as function of 
the state at the previous time and the weight the state at the current time should obtain, so instead of 
working with $W(\bX,I)$ we choose $\bX(W,I)$, which 
opens up a whole new range of efficient particle filters in high dimensional systems. 

The EWPF and the IEWPF are by construction particle filters that are not degenerate in high-dimensional systems 
 and that do not rely on localisation.
However, it is easy to see that both filters are biased, or inconsistent. In the limit of an infinite number of particles
the target-weight constructions will prevent the schemes to converge to the full posterior pdf.
The schemes are only of interest when the ensemble size is limited. 
As long as the bias from the target-weight construction is
smaller than the Monte-Carlo error this bias is of no direct consequence. 
It will be clear that the number of possible methods
that have this property is huge, and much more research is needed to explore the best possibilities.


\section{Transportation Particle Filters}

In resampling particle filters the prior particles are first weighted to represent the posterior and then transformed to 
unweighted particles simply by duplicating high-weight particles and abandoning low-weight particles. 
In transformation particle filters one tries to find a transformation that moves particles from the prior
to particles of the posterior in a deterministic manner. A related approach, which uses random transformation steps, 
is based on tempering the likelihood, which we also discuss in this section.

\subsection{One-step transportation}
In one-step transportation one tries to transform samples from the prior into samples from the posterior in 
one transformation step. An example is the Ensemble Transform Particle Filter \citep[ETPF,][]{reich13}, in which the unweighted particles are linear combinations
of the weighted particles, so one writes:
\begin{equation}
\label{ensemble transform}
\bX^a = \bX^f \bD
\end{equation}
in which the matrix $\bX^f = (\bx_1^f,\cdots,\bx_N^f)$ and similar for $\bX^a$, and in which $\bD$ is a transformation 
matrix. The only conditions on $\bD$ are that $d_{ij} \geq 0$,  $\sum_i d_{ij}=1$ and $\sum_j d_{ij}=w_i N$. These three conditions
leave a lot of freedom for all $N^2$ elements of $D$, and a useful way to determine them is to ensure minimal
overall movement in state space of the particles from prior to posterior. This leads to an optimal transportation 
problem and is typically solved by minimizing a cost function that penalises movement of particles. 

We can see immediately that this method will not work when the weights are degenerate as the solution will be 
degenerate and all particles have no other choice than move to the prior particle with weight (close to) one. 
However, the strength of this filter is that it allows for localisation in a very natural way by making the weights,
and hence the matrix $\bD$, space dependent. The method will be discussed in more detail in Section \ref{sec:Loc} on localisation.
Here we provide the basic algorithm in Algorithm \ref{alg:ETPF}.

\begin{algorithm}[ht]
	\caption{ETPF \label{alg:ETPF}}
	\begin{algorithmic}[0]
		\State $w_i=p(\by | \bx_i^f)$ 
                 \State $J(T) \gets \sum_{i,j}^{N} t_{ij} ||\bx_i^f-\bx_j^f||^2$  
	         \State Solve $\min_{T} J(T)$ with $t_{ij} \geq 0$ ,  $\sum_i^{N} t_{ij}=\frac{1}{N}$ and  $\sum_j^{N}=w_i$  
	         \State $ \bx_j^a \gets N \sum_i \bx_i^f t_{ij}^* $
	\end{algorithmic}
\end{algorithm}

The ETPF provides a direct map from prior to posterior particles without explicitly constructing a transformation map. 
An alternative approach has been suggested in \cite{sr:marzouk11}, where an approximate transportation map $\tilde \bT$ is constructed 
such that $\tilde \bT$ belongs to certain family of maps and $\tilde \bT$ is chosen such that the Kullbeck-Leibler 
divergence between the pdf generated by $\tilde \bT$ and the posterior pdf is
minimized. See \cite{Spantini2017} for an 
efficient implementation in the context of filtering and smoothing for low-dimensional systems.

\subsection{Tempering of the likelihood}
\begin{figure}[h]
	\centering\includegraphics[width=1\linewidth]{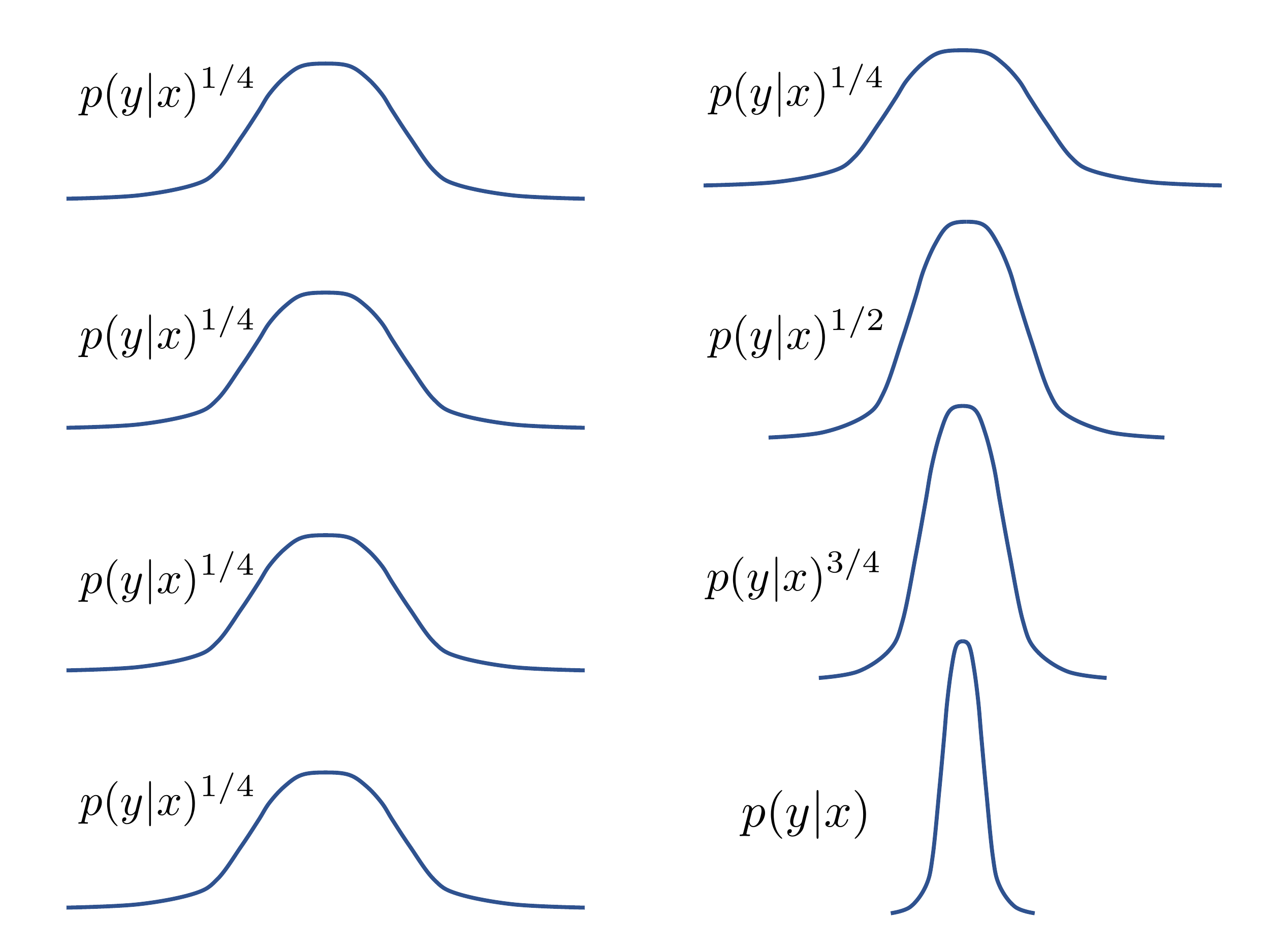}   
	\caption{Tempering. The left hand side shows the tempered likelihood functions
	used in every iteration of the tempering scheme, so every particle filter update. 
	We have chosen $\gamma_i=1/4$ in this example. The right hand side illustrates 
	how the full likelihood is build up during the tempering process.
	} 
	\label{fig:tempering-PF}
\end{figure}

Instead of trying to transform the particles from the prior to particles from the posterior in one step
one can also make this a smoother transition. In tempering (\cite{Neal1996}, see also \cite{DelMoral2006} 
and  \cite{Beskos2014}) one factorises the likelihood as follows:
\begin{equation}
p(\by|\bx) = p(\by|\bx)^{\gamma_1}... p(\by|\bx)^{\gamma_m}
\end{equation}
with $0<\gamma_i<1$ and ensuring that the sum of the $\gamma$'s is equal to 1.
Then the weighting of the particle filter is first done with the first factor, so
\begin{equation}
p_1(\bx|\by) = \frac{p(\by|\bx)^{\gamma_1}}{p(\by)^{\gamma_1}}p(\bx)
\end{equation}
The reason for this is that the likelihood is much less peaked, and hence the degeneracy
can be avoided when $\gamma_1$ is small enough. 
Figure \ref{fig:tempering-PF} illustrates the basic idea.

The particles are resampled, and now the weighting is performed using the second factor,
followed by resampling, etc. In this way the scheme slowly moves all particles
towards the high-probability regions of the posterior.
Of course, after resampling several particles will be identical, 
so one needs to jitter the particles, so perturb them slightly, to regain diversity.

This jittering should be a move of the particles that preserves the posterior
pdf. It could be implemented as a Markov-Chain Monte-Carlo method with
the posterior as the target density, e.g. exploring resample-move strategies, see
e.g. \cite{Doucet2001}. A problem is, however, that in sequential filtering we only have a representation
of the posterior density in terms of the present particles, and this representation is 
very poor due to the small number of particles. Possible avenues are to fit a pdf of a certain
shape to the present particles, e.g. a Gaussian mixture model, and use that as target
density. 

A problem in the geosciences is that this posterior fit needs to preserve the delicate balances 
between the model variables that are present in each particle, and an extra complication is that these
balances can even be nonlinear. Also the transition kernel of the Markov Chain
should somehow preserve these balances. An example of its use in the geosciences 
is the Multiple Data Assimilations (MDA) method of \cite{Emerick2013}, in which
the intermediate pdf's are assumed to be Gaussian. See also \cite{Evensen2018} for 
a comparison of this method to other iterative implementations of the Ensemble Kalman Filter/Smoother.

If, however, one allows for model error in the model equations, the following scheme proposed by \cite{Beskos2014}
does not have this problem. In that case the prior at observation time can be written as (see equation
\ref{eq:prior}):
\begin{equation}
p(\bx^{n}) \approx \frac{1}{N} \sum_{i=1}^N p(\bx^n|\bx_i^{n-1})
\end{equation}
in which we assume equal-weight particles at time $n-1$ for ease of presentation.
In this case the MCMC method that has the posterior as invariant density is easy to find
as the transition densities defined above, followed by an accept/reject step. 

When several model time steps are performed between observation times one can also perform 
tempering in the time domain, as explored in \cite{VanLeeuwen2003b}
and \cite{VanLeeuwen2009} in the Guided Particle Filter. The idea is to assimilate the observations ahead of time,
with using as likelihood $p(\by^*|\bx^{m})^{\gamma})$, in which $\by^*$ is taken equal to the value $\by^n$, and $\gamma<<1$.
Here $m<n$ is the present time of the model. This is then followed by a resampling step.
The procedure can be followed over
several time instances during the forward integration of the particles, increasing $\gamma_i$ each time.
At the observation time $\gamma=1$ is used.
This will force the particles towards the observations and does not need extra jittering because each particle
will see a different model noise realisation $\bbeta$ in the model integration after the resampling steps.

Of course one has to compensate for the fact that the transition density has been changed,
and the way to do that is to realise that we have used importance sampling.
Instead of sampling from $p(\bx^m|\bx_i^{m-1})$, we sample from a pdf 
$q(\bx^m|\bx_i^{m-1},\by^n) \propto p(\bx^m|\bx_i^{m-1})p(\by^n|\bx^m)^{\gamma}$,
in which $\by^*$ is equal to $\by^n$ taken at time $m$, and with larger observation uncertainty related to $\gamma$. 
This means that we have to compensate for the weights
created by this sampling, so we need to introduce particle weights $w_i^m = p(\bx_i^m|\bx_i^{m-1})/q(\bx_i^m|\bx_i^{m-1},\by^*) \propto 1/p(\by^*|\bx_i^m)^{\gamma}$ at each model time step we use this scheme.

The scheme generates extra weights during the model integration, but corrects for them 
at each new time when we resample, ensuring much better positioned particles at the
actual observation time $n$. It has been used in a reduced-gravity primitive equation model in \cite{VanLeeuwen2003b},
but not in high-dimensional settings.

\subsection{Particle flow filters}
There is a recent surge in methods that dynamically move the particles in state space
from equal-weight particles representing the prior, $p(\bx)$, to equal-weight particles representing the posterior,
$p(\bx |\by)$. In other words, one seeks a differential equation 
\begin{equation} \label{ODE}
\frac{d}{ds} \bx = \bff_s(\bx)
\end{equation}
in artificial time $s\ge 0$ with the flow map defining the desired transformation. 
If the initial conditions of the differential equation (\ref{ODE}) are chosen from
a pdf $p_0(\bx)$, then the solutions follow a distribution characterized by the Liouville equation
\begin{equation} \label{Liouville}
\partial_s p_s = - \nabla_\bx \cdot (p_s \bff_s)\,.
\end{equation}
with initial condition $p_0(\bx) = p(\bx)$ and final condition $p_{s_{final}}(\bx)=p(\bx|\by)$.

Two classes of particle flow filters arise. In the first we start from the tempering approach,
such that $s_{final}=1$.
We now take the limit of more and more tempering steps by choosing $\gamma_i=1/n = \Delta s$
with $\lim_{n \to \infty}$, so $\lim_{\gamma_i \to 0} $, or $\lim_{\Delta s \to  0}$, see
\cite{conf/icassp/DaumH11,conf/fusion/DaumH13,sr:reich10}. This leads to:
\begin{eqnarray}
\lim_{\Delta s \to 0} p_{s+\Delta s}(\bx) & = & p_s(\bx) \left(\frac{p(\by|\bx)}{p(\by)}\right)^{\Delta s} \nonumber \\
& = & p_s(\bx) \exp \left[ \Delta s \left(\log p(\by|\bx)-\log p(\by) \right) \right] \nonumber \\
& \approx & p_s(\bx) \left[ 1 - \Delta s \log p(\by|\bx) - \Delta s \log p(\by) \right] 
\end{eqnarray}
Hence we find:
\begin{equation} \label{ellipticPDE}
\partial_s p_s(\bx)  = -\nabla_\bx \cdot (p_s \bff_s) = p_s(\bx) (\log p(\by|\bx) - c_s)
\end{equation}
with $c_s = \int  p_s(\bx) \log p(\by|\bx)d\bx$.  Explicit expression for $f_s$ are available for certain pdfs such as 
Gaussians and Gaussian mixtures \citep{sr:reich11}. These particle flow filters can be viewed as a continuous limit of 
the tempering methods described in the previous subsection, avoiding the need for resampling and jittering. 
Note that the elliptic partial differential equation (\ref{ellipticPDE}) does not determine $f_s$ uniquely. Optimal choices
in the sense of minimizing the $L_2(p_s)$--norm of $f_s$ lead to the theory of optimal transportation, 
see \cite{Villani2008} and \cite{reichcotter15}.

\begin{figure}[h]
	\centering\includegraphics[width=1\linewidth]{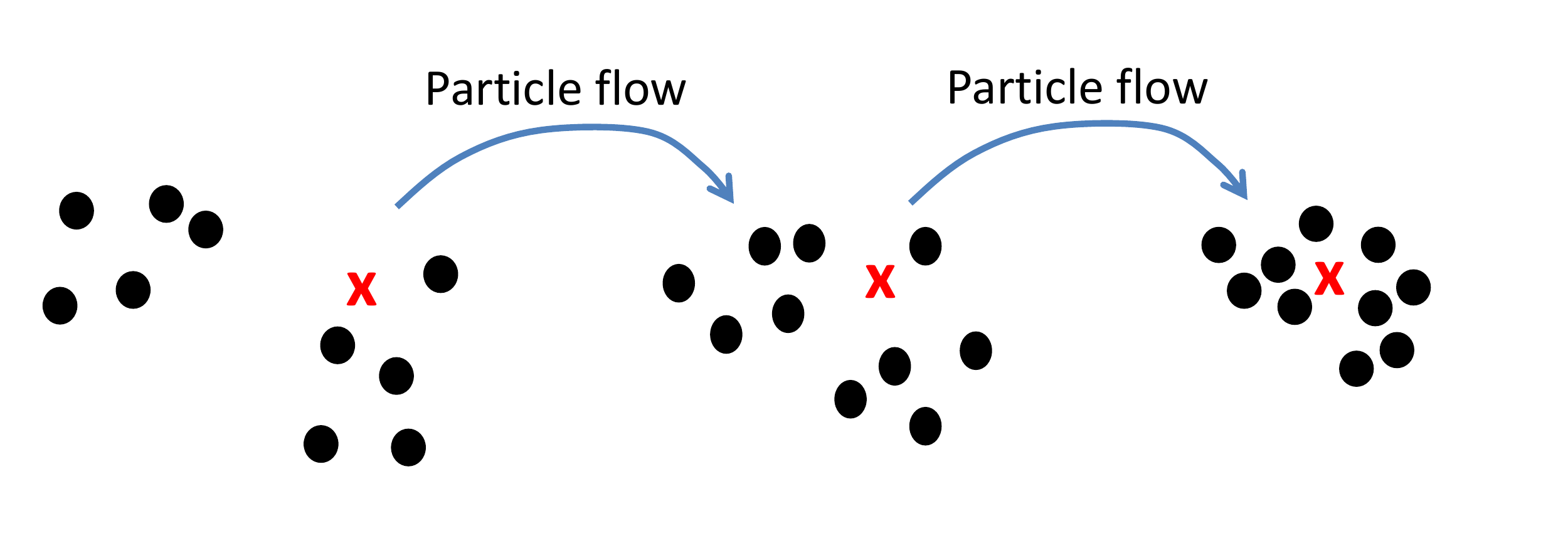}   
	\caption{A typical particle flow filter. Left: the prior particles (dots), with one observation,
	denoted with the red cross. Middle: the particles have moved over several artificial time steps
	towards the posterior. Note that the weights do not change. Right: the posterior particles
	after convergence of the filter, sampling the posterior directly.} 
	\label{fig:Particle-Flow-PF}
\end{figure}

Figure \ref{fig:Particle-Flow-PF} shows the basic idea behind particle flow filters.

Alternatively, one can explore ideas from Markov-Chain Monte Carlo (MCMC). 
One MCMC method that generates samples from the posterior is the Langevin Monte-Carlo sampling,
in which a sequence of samples is generated by
\begin{equation}
x^{j+1} = x^j - \Delta s \nabla_\bx \log p(\bx|\by) + \sqrt{2\Delta s} \beta^j
\end{equation}
in which $\beta^j$ a random forcing term drawn from $N(0,\bI)$.
One can show that in the limit of $j \to \infty$ these samples will be samples from the posterior.
The corresponding Fokker-Planck equation for this stochastic PDE reads
(see, for example, \cite{reichcotter15}):
\begin{align*}
\partial_s p_s &= \nabla_\bx \cdot \left(p_s \nabla_\bx (-\log p(\bx|\by)) \right) +  \nabla_\bx \cdot \nabla_\bx p_s \\
&= -\nabla_\bx \cdot \left(p_s \left\{ \nabla_\bx.\log p(\bx|\by) - \nabla_\bx \log p_s \right\} \right)
\end{align*}
This equation corresponds to the deterministic PDE (\ref{ODE})
in which $\bff_s(\bx)$ is given by:
\begin{equation} \label{VV-FP}
\bff_s(\bx) := \nabla_\bx \log p(\bx|\by) - \nabla_\bx \log p_s(\bx) = -\nabla_\bx \log \frac{p_s(\bx)}{p(\bx | \by)}
\end{equation}

Many other choices are possible that use
\begin{equation}
\lim_{s\to \infty} p_s = p(\bx | \by)
\end{equation}
in (\ref{Liouville}). An alternative approach, called Stein variational descent, 
has recently been proposed by \cite{Liu2016}. Stein variational descent can be viewed as a numerical approximation
to a particle flow (\ref{ODE}) with vector field
\begin{equation} \label{VV-SVD}
\bff_s(\bx) := p_s \left( \nabla_\bx \log p(\bx|by) - \nabla_\bx \log p_s(\bx) \right)
\end{equation}
\citep{LLN18}. We come back to this method below.

In general, to use any of these methods we need to be able to evaluate $p_s(\bx_i)$, which is typically unknown
as we only know the particle representation of $p_s(\bx)$.
One way to solve this issue is to explore kernel embedding.
A numerical implementation of the two formulations (\ref{VV-FP}) and (\ref{VV-SVD}) 
can be based on a reproducing-kernel Hilbert space (RKHS) $\cal{F}$ with reproducing kernel $K(.,.)$, typically taken as a Gaussian. 
In the sequel, we will therefore assume that the kernel is symmetric $K({\bx,\bz}) = K(\bz,\bx)$.
The inner product $\langle g,f\rangle_{\cal F}$ in $\cal{F}$ satisfies the reproducing property
\begin{equation}
g(\bx) = \langle K(\bx,\cdot),g \rangle_{\cal F}\,.
\end{equation}
A computational approximation to (\ref{VV-FP}) can now be obtained as follows \citep{R90,DM90}.
One approximates the pdf $p_s$ by
\begin{equation}
p_s(\bx) = \frac{1}{N} \sum_{j=1}^N K(\bx_j,\bx)\,,
\label{eq:kernel-approx}
\end{equation}
the vector field $\bff_s$ by
\begin{equation}
\bff_s(\bx) = \frac{\sum_{j=1}^N K(\bx_j,\bx) \bu_s^j}{p_s(\bx)}\,,
\end{equation}
and the $N$ particles $\bx_j$ move under the differential equations
\begin{equation} \label{ODEx}
\frac{d}{ds} \bx_j= \bu_s^j\,.
\end{equation}
Since the drift term (\ref{VV-FP}) gives 
rise to a gradient flow in the space of pdfs with respect to the 
Kullback--Leibler divergence ${\rm KL} = {\rm KL}(p_s||p(\cdot|\by))$ 
between $p_s$ and the posterior pdf \citep{reichcotter15}, it is natural to introduce 
the following particle approximation of the Kullback--Leibler divergence:
\begin{equation}
{\cal V}(\{\bx_l\}) := \left \langle p_s,\log \frac{p_s}{p(\cdot |\by)}\right \rangle_{\cal F}\,.
\label{eq:potential}
\end{equation}
in the RKHS $\cal{F}$ and to set
\begin{equation}
\bu_s^j := - N \nabla_{\bx_j} {\cal V}(\{\bx_l\})
\end{equation}
in (\ref{ODEx}), which leads to a gradient flow in the particles $\{\bx_l\}$ minimising ${\cal V}$.
Details on the numerical implementation of this approach can be found in \cite{sr:PR19}.

The above formulation restricts the pdf $p_s$, and hence the prior and the posterior,
to be of the form (\ref{eq:kernel-approx}).
Alternatively, one can embed the vector field of the flow in an appropriate reproducing kernel Hilbert space
and not the density itself.
With that we can derive a practical implementation of the Stein variational formulation (\ref{VV-SVD}) as follows.
First, note that the change in KL due to the flow field $\bff_s$ can easily be found as:
\begin{eqnarray}
dKL & = & \lim_{\epsilon \to 0} \frac{{\rm KL}(p_{s+\epsilon})-{\rm KL}(p_s)}{\epsilon}                 \nonumber \\
& =  & - \int p_s(\bx) \left[ \bff_s(\bx)^T\nabla_\bx \log p(\bx|\by) + \nabla_\bx \cdot \bff_s(\bx)\right] d\bx . \nonumber \\
&  = &  \left\langle \nabla {\rm KL}, \bff_s \right\rangle_{\cal F}.
\label{eq:variational}
\end{eqnarray}
where $\nabla {\rm KL}$ is the gradient of ${\rm KL}$, the maximal functional derivative of ${\rm KL}$ at every state vector $\bx$ in the RKHS. Note that $\cal F$ here is different from the Hilbert space used earlier. 
Maximising this change in KL as function of the flow field $\bff_s$ is not trivial in general.
However, with the reproducing kernel property of $\bff_s$ we have
\begin{equation}
\bff_s(\bx) = \left \langle {\cal{K}} (\cdot, \bx) , \bff_s (\cdot) \right \rangle
\end{equation}
in which $\cal{K}$ is a vector-valued kernel, typically taken as ${\cal{K}} = \bI K$. Using this in  (\ref{eq:variational}), 
the gradient of the {\rm KL} divergence is found as
\begin{equation} \label{VD_KL}
\nabla {\rm KL} (\bx) = -
\int p_s(\bz) \left[ K(\bz,\bx)\nabla_\bz  \log p(\bz | \by)+ \nabla_\bz K(\bz,\bx)\right] d\bz\,.
\end{equation}
The important point is that this gradient is independent from $\bff_s$.
One now chooses $\bff_s$ along this direction, which gives the steepest descent, as
\begin{equation}
\bff_s(\bx) = - \epsilon \nabla {\rm KL} (\bx)
\end{equation}
Finally, one replaces the integral in (\ref{VD_KL}) by its empirical approximation, to obtain
\begin{equation} \label{SVD}
\bff_s(\bx_j) = \epsilon \frac{1}{N} \sum_{l=1}^N \left[K(\bx_l,\bx_j)\nabla_\bx \log p(\bx_l | \by) + \nabla_{\bx} K(\bx_l,\bx_j) \right]
 \end{equation}
 for the dynamics (\ref{ODE}) of the $N$ particles $\bx_j$.
 
The intuition behind Stein variational descent is that the first term in (\ref{SVD}) pulls the particles
 towards the mode of the posterior, while the second term acts as a repulsive force that allows
 for particle diversity. \cite{Liu2016} derived this formulation for a steady-state problem, and \cite{Pulido2018}
 have extended the method to sequential particle filters.
 The scheme is given in Algorithm \ref{alg:algo}.
 
 \begin{algorithm}
\caption{Mapping Particle Filter}\label{alg:algo}
\begin{algorithmic}
  \For{$j=1,N$}
    \State $\bx^{k,0}_j  \gets f(\bx^{k-1}_j,\beta^{k})$
    \EndFor
    \State $i=1$
\Repeat  
   \For{$j=1,N$}
   \State $ \nabla KL(\bx) \gets - \frac{1}{N} \sum_{l=1}^{N} \left[ K(\bx^{k,i-1}_l, \bx ) \nabla \log p(\bx^{k,i-1}_l | \by)\right. $ 
   \State $\qquad \left. +  \nabla_x K(\bx^{k,i-1}_l, \bx)\right] $
   \State $\bx_j^{k,i} \gets \bx_j^{k,i-1}- \epsilon  \nabla KL(\bx_j^{k,i-1})$
   \EndFor
   \State $ i \gets i+1$
\Until Stopping criterion met
\end{algorithmic}
\end{algorithm}
 
The free parameter of these methods is the reproducing kernel $K(.,.)$,
which needs to be chosen such that the particles sample the posterior and
that physical (and potentially other) balances are retained. 
One also needs to select a proper time stepping scheme, typically chosen as a forward Euler scheme with variable 
time step $\epsilon$, which can now be viewed as the step length in a gradient descent optimisation algorithm.

\subsection{Discussion}
Viewing particle filters as a transportation problem from equal-weight particles of the prior 
to equal-weight particles of the posterior has led to an interesting set of filters. 
None of them have been implemented yet in high-dimensional settings, but some of them
are ready to do so. 
The strong involvement of the machine learning community in problems 
of this kind also suggests rapid progress here.
Finally we mention that the equal-weight particle filters from section 2 can be viewed as
one-step transportation filters that explore the proposal density freedom, and in fact
transform equal-weight prior particles at time $n-1$ to equal-weight posterior particles at observation time $n$.

\section{Localisation in Particle Filters}
\label{sec:Loc}
Localisation is a standard technique in Ensemble Kalman filtering to increase the rank of the
ensemble perturbation matrix, allowing for more observations to be assimilated, and to suppress
spurious correlations where real correlations are very small, but ensemble correlations are 
larger because of sampling noise. Localisation limits the influence of each observation to a localisation area that is much smaller than the full model domain.
This idea can easily be incorporated when calculating the particle weights locally, as pioneered by 
\cite{Bengtsson03}, and \cite{VanLeeuwen2003b}, and used in a high-dimensional parameter estimation problem
in \cite{Vossepoel2006}. The difficulty, as we shall see, lies in the resampling step:
how does one generate 'smooth' global particles from locally resampled particles.
Smooth is not well defined here, but it is related to the particles having realistic
physical relations (balances) between the model variables. For example, if geostrophic balance 
is dominant, the resampling procedure should not generate particles that are completely
out of geostrophic balance as that would lead to spurious adjustment processes via spurious
gravity waves. Up to now localisation is mainly used in connection with the standard particle
filter, while more advanced proposals, apart from the optimal proposal, have not been explored.
\cite{Farchi2018} provide an excellent review of localisation in particle filtering,
treating a subset of the methods presented here, but including interesting extensions of 
the methods they describe.

\begin{figure}[h]
	\centering\includegraphics[width=1\linewidth]{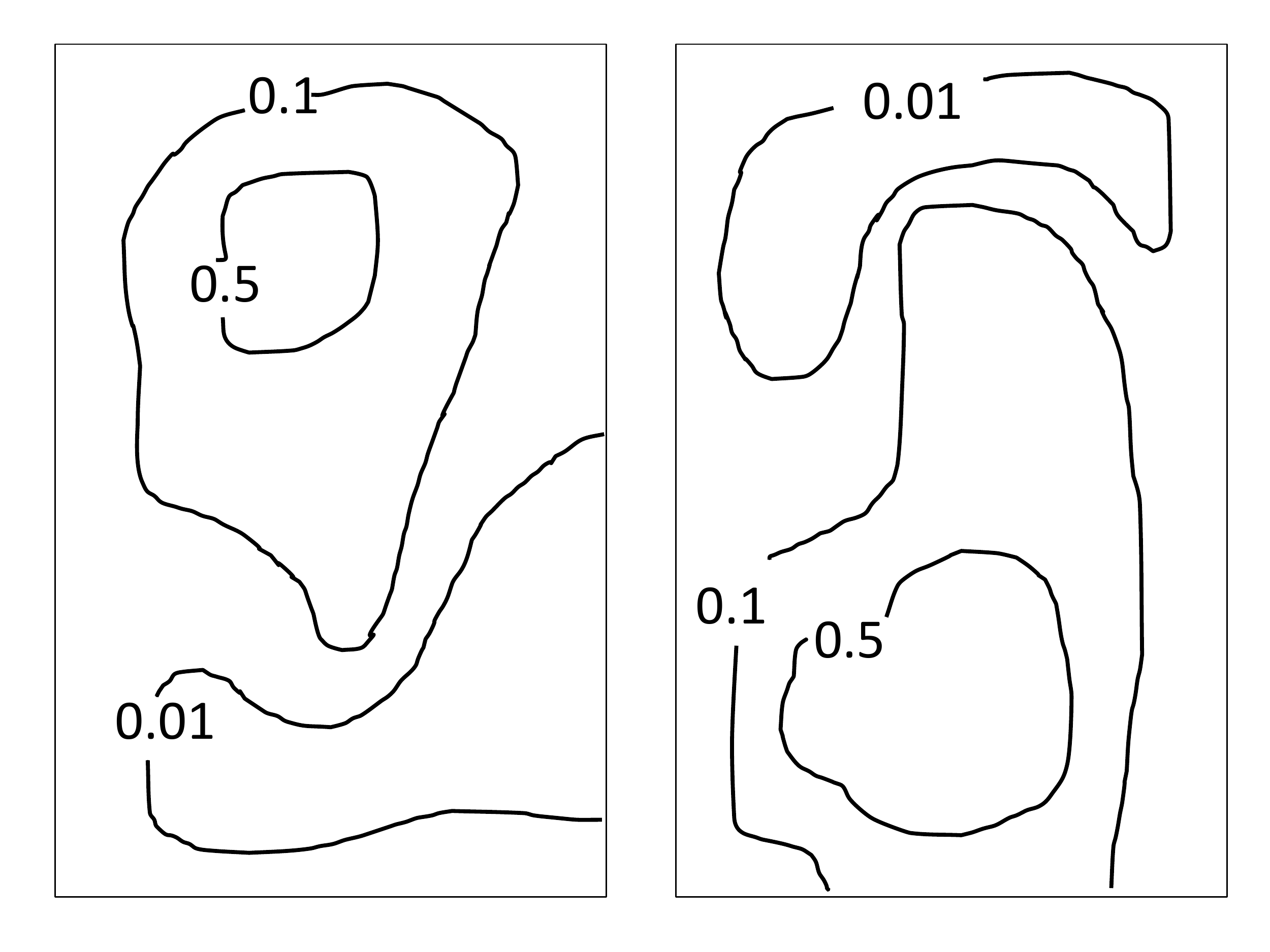}   
	\caption{Illustration of a possible local weight distribution in a two-dimensional domain,
	for two different particles. The particle on the left is close to observations in the central upper part
	of the domain, leading to high weights there, while the particle on the right is closer to observations
	in the central lower part of the domain, and hence higher weights there. } 
	\label{fig:Local-weighting}
\end{figure}

The formal way localisation can be introduced in particle filtering is as follows.
Let us denote the state at grid point $k$ as $\bx^k$. Hence in contrast to other sections a superscript here denotes 
not the time index, but the grid point. Note that in geoscience applications
each grid point typically has several model variables, so $\bx^k$ is a vector in general.
Physically it makes sense to assume
that the posterior of the state at this grid point depends only on a subset
of the observations. Let us denote that subset as $\by^{[k]}$. We can then write:
\begin{equation}
\label{eq:local-post}
p(\bx^k|\by) \approx p(\bx^k | \by^{[k]})
\end{equation}
In turn, these observations do not depend on the whole state vector but
only on part of it, denoted by $\bx^{(k)}$:
\begin{equation}
p(\by^{[k]}|\bx) = p(\by^{[k]}|\bx^{(k)})
\end{equation}
Introduce the notation $\bx^{(k)\setminus k}$ to denote all those grid points in that part of the state vector
excluding grid point $k$. Then we can rewrite the above as an integral over the 
joint pdf:
\begin{equation}
p(\bx^k | \by^{[k]})= \int p(\bx^{(k)}|\by^{[k]}) \;d\bx^{(k)\setminus k}
\end{equation}
Exploring Bayes Theorem we find:
\begin{eqnarray}
p(\bx^{(k)}|\by^{[k]}) &=& \frac{p(\by^{[k]}|\bx^{(k)})}{p(\by^{[k]})}
p(\bx^{(k)}) \nonumber \\
& \approx & \frac{1}{N} \sum_i^N \frac{p(\by^{[k]}|\bx_i^{(k)})}{p(\by^{[k]})} 
\delta(\bx^{(k)} - \bx_i^{(k)}) \nonumber \\
 & = & \sum_i^N w_i^{(k)} \delta(\bx^{(k)} - \bx_i^{(k)})
\end{eqnarray}
Taken together, this shows that
\begin{equation}
p(\bx^k| \by^{[k]}) \approx \sum_i^N w_i^{(k)} \delta(\bx^{k} - \bx_i^{k})
\end{equation}
The weights $w^k_i$ thus depend only on the local observations $\by^{[k]}$ and
the local prior particles $\bx^{(k)}_i$, so that the variance
of the weights will be much smaller.
Figure \ref{fig:Local-weighting} illustrates how this local weighting could look for two different particles.

The approximation \eqref{eq:local-post} is not unrealistic:
a temperature observation in New York is not expected to change our pdf of
the temperature in London at the moment of the observation. There will be,
of course, an effect at later times, but that is not relevant here.
The same assumption underlies the use of localisation in Ensemble Kalman Filters,
and in variational methods when the background error covariance is
constructed. 

However, mathematically it does not follow from the assumption
that under the prior the values of the state at grid points separated by
more than a certain distance are independent. There can be an indirect flow
of information from observations far apart over observations between 
neigboring grid points. In Ensemble Kalman Filters, the Kalman gain is
generally a dense matrix even if $\bH \bP^b \bH^T +  \bR$ is sparse, because
its inverse $(\bH \bP^b \bH^T +  \bR)^{-1}$ can be dense. On the other hand,
if $\bH \bP^b \bH^T +  \bR$ is diagonally dominant, then often its inverse is too.

Repeating the localisation procedure for all grid points, we obtain all marginals
of the posterior pdf. However, because the weights $w^{(k)}_i$ change
from one grid point to the next, it is non-trivial to obtain a consistent
posterior for pairs of state values $(\bx^k,\bx^\ell)$ (and similarly for
triplets etc.). This can easily be seen using Fig.\ 5: we would like to retain the
left particle in the central upper half of the domain, and abandon elsewhere.
That would mean that where ever it is abandoned we need to replace it
with another particle, perhaps partly with the particle in the right part of the
figure. At the boundary between particles a discontinuity will exist,
which will lead to unphysical behaviour when this new particle is propagated
forward in time.

This means that to obtain global
particles that can be forwarded with the model equations  
one would need to somehow smoothly glue different particles together. This is a
major problem 
and has hampered localisation in particle filtering since the early 2000's. 
However, recently clever smoothing schemes have been constructed that seem
to work well in high-dimensional geophysical applications. We will report on those
below.

Another issue is that the localisation area cannot be too large to avoid filter collapse. 
As a rule of thumb, when there are more than say 10 independent observations inside 
a local area, the particle filter will still tend to be degenerate for the number of $O(10-1000)$ particles one can typically
afford. This means that when
the observation density is high the localisation areas have to become unphysically small, 
or observations have to be discarded. This issue might be solved using tempering techniques
as discussed earlier, but is often avoided by artificially enforcing a minimal weight of the particles,
or by changing the observations, for instance by projecting them on a lower dimensional
space favoured by the prior.

Setting a minimal weight or projecting observations to a lower dimensional space favoured by the prior
has as consequence that not all information will be extracted from the
observations, as observations that are very different from the existing particles will be largely ignored. 
This is not directly equivalent to the standard quality control measures used by operational
weather forecasting centres, in which observations that are a few standard deviations away from the forecast are ignored.
The issue here is that a distance of less then one standard deviation for a few observations can 
already lead to weight collapse, and artificially setting minimum values for the weights avoids that.

\subsection{Localisation based on resampling}

Several localisation schemes have been proposed and discussed in the review by \cite{VanLeeuwen2009} and those will not be repeated here.
The most obvious thing to do is to weight and resample locally, and somehow glue the resampled particles together
via averaging at the edges between resampled local particles \citep{VanLeeuwen2003b}.
In the following, several schemes in this category are discussed.

\subsubsection{The Localized Particle Filter}
Recently, \cite{Penny2016} used this idea with more extensive averaging, and their scheme runs as follows.
First, for each grid point $j$ the observations close to that grid point are found and the weight of each particle $i$
is calculated based on the likelihood of only those observations:
\begin{equation}
w_{i,j}= \frac{p(\by_j | \bx_{i,j})}{\sum_{k=1}^N p(\by_j | \bx_{k,j})}
\end{equation}
in which $\by_j$ denotes the set of observations within the localisation area. Note the change of notation 
from the previous section, related to the explicit use of the particle index in all the following.
This is followed by resampling via Stochastic Universal Resampling to provide ensemble members $\bx^a_{i,j}$
with $i=1,...,N$ for each grid point $j$.

\cite{Farchi2018} extended this methodology by updating blocks of grid points locally, and introduce
a smoothing operator in the weights (similar to \cite{Poterjoy2016a}), as:
\begin{equation}
w_{i,j}= \frac{\sum_{k=1}^{N_{j}}G(d_{j,k}/h) (p(\by_{k} | \bx_{i,k})}{\sum_{m=1}^N \sum_{k=1}^{N_{j}}G(d_{j,k}/h) p(\by_{k} | \bx_{m,k})}
\end{equation}
in which $G(..)$ is a distance weighting function, e.g. a Gaussian or an approximation of that, 
$d_{j,k}$ is the distance between grid points $j$ and $k$, for each observation $\by_{k}$ 
at grid point $k$ in the neighbourhood of grid point $j$. The parameter $h$ is a distance radius, 
another tuning parameter. This formulation can be used for each grid
point $j$, but also for each block of grid points $j$. They note that $G$ can also be a Gaussian 
of a Gaussian, such that it works directly on $-\log p(\by_{k}|\bx_{i,k})$.

As mentioned before, the issue is that two neighbouring grid points can have different sets of particles, and
smoothing is needed to ensure that the posterior ensemble consists of smooth particles.
This smoothing is performed by \cite{Penny2016} for each grid point $j$ for each particle $i$ by averaging over the $Np$ neighbouring points
within the localisation area around grid point $j$:
\begin{equation}
x^a_{i,j}= \frac{1}{2}x^a_{i,j} + \frac{1}{2Np} \sum_{k=1}^{N_p} x^a_{i,j_k}
\end{equation}
in which $j_k$ for $k=1,...,N_p$ denotes the grid point index for those points in the localisation area around grid point $j$.
The resampling via Stochastic Universal Resampling is done such that the weights are sorted before the resampling, so that
high-weight particles are joined up to reduce spurious gradients.

\cite{Farchi2018} also suggest to smooth this operation, as follows:
\begin{equation}
x^a_{i,j}= \alpha x^a_{i,j} + (1-\alpha)  \sum_{k=1}^{N_p} G(d_{j,{j_k}}/h) x^a_{i,j_k}
\end{equation}
with $\alpha$ a tuning parameter. Note that choosing $\alpha=1/2$ and
$G(d_{j,j_k}/h) = 1/N_p$ we recover the scheme by \cite{Penny2016}.

While these schemes have been shown to solve the degeneracy problem in intermediate dimensional systems 
with fixed balances, like the barotropic vorticity model, it is unclear 
how they will perform in complex systems such as
the atmosphere in which fronts can easily be smoothed out, and nonlinear balances broken, see e.g. the discussion
in \cite{VanLeeuwen2009}.  

\subsubsection{The Local Particle Filter}
A different scheme that involves a very careful process of 
ensuring smooth posterior particles and retaining nonlinear relations has recently been proposed by \cite{Poterjoy2016a}. 
An important difference with the state-space localisation methods discussed above is that
observations are assimilated sequentially to avoid the discontinuity issues of the state-space localisation.
This makes the algorithm non-parallel, so slower than the state-space localisation methods,
but \cite{Farchi2018} demonstrate that a lower root-mean square error (RMSE) can be achieved. 

The scheme proceeds as follows.
First, adapted weights are calculated for the first element $y_1$ of the observation vector, as
\begin{equation}
\tilde{w}_i = \alpha p(y_1 | \bx_i)+ 1-\alpha 
\end{equation}
These weights are then normalised by their sum $\tilde{W}$.
Then the ensemble is resampled according to these normalised weights to form particles $\bx_{k_i}$.

The scalar $\alpha$ is an important parameter is this scheme, with $\alpha=1$ leading to standard weighting, 
and $\alpha=0$ leading to all weights being equal to 1 (before normalisation).
Its importance lies in the fact that the weights are always larger than $1-\alpha$, so even
a value close to 1, say $\alpha=0.99$, leads to a minimum weight of $0.01$ that might seem 
small, but it means that particles that are more
then 1.7 observational standard deviations away from the observations have their weights cut off to a value close to  $1-\alpha$.
This limits the influence the observation can have on the ensemble. 
Furthermore, the influence of $\alpha$ does depend on the size of the observational error, which
is perhaps not what one would like.
It is included to avoid loosing any particle.

Now the following is done for each grid point $j$. 
For each member $i$ a weight is calculated as
\begin{equation}
\tilde{\omega}_{i} = \alpha \rho(1,j,r) p(y_1 |\bx_i)  + 1- \alpha \rho(1,j,r)
\end{equation}
in which $\rho(..)$ is the localisation function with localisation radius $r$.
These weights are normalised with their sum over the particles, so a normalised weight $\omega_i$ for this grid point {is obtained.
Note, again, the role played by $\alpha$.
Then the posterior mean for this observation at this grid point is calculated as
\begin{equation}
\bar{\bx}_j=\sum_{i=1}^N \omega_i \bx_{i,j}
\end{equation}
in which $\bx_{i,j}$ is the state at grid point $j$ of particle $i$.
Next a number of scalars are calculated that ensure smooth posterior fields \citep{Poterjoy2016a}
as detailed in Algorithm \ref{alg:LPF}.

The final estimate becomes:
\begin{equation}
\bx^a_{i,j}  = \bar{\bx}_j+r_{1j}(\bx_{k_i,j}-\bar{\bx}_j) + r_{2j}(\bx_{i,j}-\bar{\bx}_j)
\end{equation}
where $k_i$ is the index of the $i$'s sampled particle.
This procedure is followed for each grid point so that at the end an updated set of particles is obtained
that have incorporated the first observation. 
As a next step the whole process is repeated for the next observation, with the small change that
$\tilde{\omega}_i$ is multiplied by $\tilde{\omega}_i$ from the previous observation, until all observations have been 
assimilated. In this way, the full weight of all observations is accumulated in the algorithm.
Now the importance of $\alpha$ comes to full light: without $\alpha$ the ensemble would collapse
because the $\tilde{\omega}$'s would be degenerate when observations are accumulated.

The final estimate shows that each particle at grid point $j$ is the posterior mean at that point
plus a contribution from the deviation of the posterior resampled particle from that mean
and a contribution from the deviation of the prior particle from that mean. 
So each particle is a mixture of posterior and prior particles,
and departures from the prior are suppressed.
When $\alpha=1$, so for a full particle filter, we find for grid points at the observation location,
for which $\rho(1,j,r)=1$, that $c_j=0$, so $r_{2j}=0$, and $r_{1j} \approx 1$, so
indeed the scheme gives back the full particle filter. 
The basic elements of the scheme are depicted in Algorithm \ref{alg:LPF}.

\begin{algorithm}[!ht]
	\caption{Local Particle Filter \label{alg:LPF}}
	\begin{algorithmic}[0]
	        \For {Each observation $l$}
		        \For {Each particle $i$}
			        \State $\tilde{w_i} \gets \alpha p(y_l | x_i) +1 - \alpha$  
			     \EndFor
			     \State $\tilde{W} \gets \sum \tilde{w_i}$
			\State Resample $\bx_{k_i}$ 
		        \For {Each grid point $j$} 
		        \For {Each particle $i$}
			    \State $\omega_i \gets \alpha \rho(l,j,r) p(y_l | x_i) +1 - \alpha \rho(l,j,r)$ 	
		        \EndFor		
		         \State $\overline{x} \gets \sum \omega_i x_{i,j}   $
			 \State $\sigma^2 \gets \sum \omega_i (x_{i,j}-\overline{x})^2 $ 
			 \State $c \gets \frac{N(1-\alpha\rho(x_j,\by_l,r))}{\alpha\rho(x_j,\by_l,r) \tilde{W}}$ 
			 \State  $r_{1} \gets \sqrt{\frac{\sigma_j^2}{\frac{1}{N-1}\sum_{i=1}^N(x_{k_i,j}-\bar{x}+c(x_{i,j}-\bar{x}))^2}}$ 
			   \State $r_{2} \gets c r_{1}$  
		            \For {Each particle $i$}
			         \State $x^a_{i,j}  \gets \bar{x}+r_{1}(x_{k_i,j}-\bar{x}) + r_{2}(x_{i,j}-\bar{x})$ 
			   \EndFor
			\EndFor
		\EndFor
	\end{algorithmic}
\end{algorithm}

At grid points between observations it can be shown that the particles have the correct first and 
second order moments, but higher-order moments are not conserved.
(\cite{Farchi2018} generate a scheme that is quite similar, but they ensure 
correct first and second moment by exploring the localised covariances between
observed and unobserved grid points directly in a regression step.)
To remedy this a probabilistic correction is applied at each grid point, as follows.
The prior particles are dressed by Gaussians with width 1 and weighted by 
the likelihood weights to generate the correct posterior pdf.
The posterior particles are dressed in the same way, each with weight $1/N$.
Then, the cumulative density functions (cdf's) for the two densities are calculated using a trapezoidal rule integration.
A cubic spline is used to find the prior cdf values at each prior particle $i$, denoted by $cdf_i$.
Then a cubic spline is fitted to the other cdf, and the posterior particle $i$ is found as the inverse 
of its cdf at value $cdf_i$. See \cite{Poterjoy2016a} for details. The result of this procedure is that
higher-order moments are brought back into the ensemble between observation points.

This scheme, although rather complicated, is one of the two local particle filter scheme that has been
applied to a high-dimensional geophysical system based on primitive equations in \cite{Poterjoy2016b}.
The other is the Localised Adaptive Particle Filter discussed below.
(\cite{VanLeeuwen2003b} applied a local particle filter to a high-dimensional quasi-geostrophic system, 
but that system is quite robust to sharp gradients as it does not allow for gravity waves.)

\subsubsection{The Localised Adaptive Particle Filter}
\label{sec:LAPF}

The {\em localized adaptive particle filter} (LAPF) is based on the localized
version of the ensemble 
transform (\ref{ensemble transform}) following the LETKF described in 
\cite{Hunt07a}, see also \cite{reich13}, with localization in observation space,
and resampling in the spirit of Gaussian Mixture filters \citep{Stordal11a}. 
Localization is carried out around each grid point, and a transform matrix ${\bD}$ is calculated for each
localization box. We note that, as for the LETKF, the weights given by (\ref{eq:weights})
depend continuously on the box location and the observations. 

In a first step, the observations are projected into the space spanned by the 
prior particles. As mentioned above, this will reduce the information extracted 
from the observations, but is perhaps less ad-hoc than setting a lower bound
on the weights, as for instance used in the LPF. 
The LAPF carries out local resampling using universal resampling (see e.g.\ \cite{VanLeeuwen2009}).

In a second step, a careful 
adaptive sampling is carried out in ensemble space around each of 
the $N$ temporary particles. 
This scheme runs as follows:

(a) Resampling is carried out based on a 
(radial) basis function centered at each particle. A simple case would be a
Gaussian mixture, where the covariance of each of the centered Gaussians is taken 
as a scaled version $c \bP$ of the local dynamical ensemble covariance $\bP$. 

(b) The scaling
factor $c$ is individually calculated for each box based on the local observation
minus background error statistics. For details we refer to \cite{potthast17_1}.
By this, the LAPF guarantees to obtain
a spread of the analysis ensemble which is consistent with the local dynamical
observation minus background (o-b) statistics and the observation error covariance $\bR$. Further standard
tools from the LETKF literature to control ensemble spread can be employed if needed.

(c) To obtain sufficient smoothness of the fields in physical
space, the LAPF uses $N$ global random draws to generate the resampling 
vectors around each particle in the space of ensemble coefficients. 
In combination with the fact that the LAPF draws in each box around each 
particle only -- in a globally uniform way modulated by the ensemble 
covariance $\bP$ and the factor $c$ only --, 
consistency and balance of the fields is achieved with sufficient
precision. The scheme is depicted in Algorithm \ref{alg:LAPF}.

\begin{algorithm}[!ht]
	\caption{Local Adaptive Particle Filter \label{alg:LAPF}}
	\begin{algorithmic}[0]
	        \For {Each grid point $j$, and local grid points $k$}
		        \State Project local $\by_k$ onto space  $\{H(\bx_{1,k}^n),..., H(\bx_{N,k}^n)\}$ 
		        \For {$i=1,..,N$}
			    \State $ w_i \gets p(\by_k|\bx_{i,k})$ 
			\EndFor
			\State $ \bw \gets \bw/\bw^T\mathbf{1}$
			\State Resample 
		\EndFor
		\State $\bP \gets Localized(\bX \bX^T) $
		\State $c < 1$ (depends on o-b statistics, see text)
		\For {$i=1,..,N$}
			\State $\bbeta \sim N(0,c\bP)$
			\State $ \bx_i \gets \bx_i + \bbeta $
		\EndFor
	\end{algorithmic}
\end{algorithm}

The LAPF is the first particle filter that has been implemented and tested in an
operational numerical weather prediction context, and we provide a short description of 
the procedure.
The method has been implemented in the data assimilation 
system DACE (Data Assimilation Coding Environment) of Deutscher Wetterdienst 
(DWD) \cite{potthast17_1}. The DACE environment includes a {\em Local 
Ensemble Transform Kalman Filter} (LETKF) based on \cite{Hunt07a} both for 
the global ICON model system and the convection permitting COSMO model system 
of DWD, see \cite{Schraff15}), both of which are run operationally at 
DWD\footnote{since January 20, 2016 for the global ICON model with 40km global 
ensemble resolution including a 20km resolved two-way nest over Europe; and 
since March 21, 2017 for the COSMO model with 2.8km resolution over central 
Europe} and build a basis, framework and reference for the LAPF particle 
filter implementation. 

The ensemble data assimilation system is equipped with a variety of tools to control the spread of 
the ensemble, such as {\em multiplicative inflation} and {\em additive 
inflation}, {\em relaxation to prior spread} (RTPS), {\em relaxation to prior 
perturbations} (RTPP) and stochastic schemes to add spread to soil moisture 
and sea surface temperature (SST) when needed (details are described in \cite{
Schraff15}).  

Tests with the LAPF for the global ICON model 
with 40 particles of 40km global resolution have been successfully 
and stably run over a duration of one month.
Extensive tests on how many
particles form the basis for resampling in each localization box have been
carried out, the numbers vary strongly over the globe and all heights of the
atmosphere, ranging from $1$ to $N$, with relatively flat distribution. 
Diagnostics and tuning of the system 
is under development and discussed in \cite{potthast17_1}.
Results show that
the quality of the LAPF does not yet reach the scores of the operational 
global LETKF-EnVAR system, but the system runs stably and forecast scores
are about 10-15\% behind the current operational system.

\subsection{The Local Ensemble Transform Particle Filter}
\label{sec:LETPF}
This filter uses a classic sequential importance resampling particle filter 
from a set of forecast particles $\bx_i^{\rm f}$, which can be obtained employing either the standard
or the optimal proposals (or any other) and their associated importance weights $w_i^{\rm f}$. 
The particles are then resampled in a statistically consistent manner, which can be characterized
by an $N\times N$ stochastic transition matrix ${\bf D}$ with the following properties: (i) all entries
$d_{ij}$ of ${\bf D}$ are non-negative and
\begin{equation} \label{marginals}
\sum_{i=1}^N d_{ij} = 1\,, \quad \frac{1}{N} \sum_{j=1}^N d_{ij} = w_i^{\rm f}\,.
\end{equation}
Let us denote the set of all such matrices by $\mathcal{D}$. Then any ${\bf D} 
\in \mathcal{D}$ leads to a resampling scheme by randomly drawing an element 
$j^\ast \in \{1,\ldots,N\}$ according to the probability vector ${\bf p}_j 
= (p_{1j},\ldots,p_{Nj}) \in \mathbb{R}^N$ for each $j=1,\ldots,N$. 
The $j$th forecast particle $\bx_j^{\rm f}$ is then replaced by $\bx_{j^\ast}^{\rm f}$ and the new particles 
$\bx_j^n = x_{j^\ast}^{\rm f}$, $j=1,\ldots,N$, provide an equally weighted set of particles from
the posterior distribution. Note that multinomial resampling corresponds to the simple choice
\begin{equation}
d_{ij} = w_i^{\rm f}.
\end{equation}
The ensemble transform particle filter (ETPF) \citep{reich13,reichcotter15} 
is based on the particular choice $\widehat{\bf D} \in {\cal D}$ that minimizes 
the expected squared Euclidian distance between forecast particles, i.e.,
\begin{equation} \label{ETPF_cost}
\widehat{\bf D} = \arg \min_{{\bf D}\in {\cal D}} 
\sum_{i,j=1}^N d_{ij} \|\bx_i^{\rm f} - \bx_j^{\rm f}\|^2\,.
\end{equation}
It has been shown under appropriate conditions that the variance of a resampling step 
based on $\widehat{\bf D}$ vanishes as $N\to \infty$ \citep{mccann95,reich13}. 
This fact is utilized by the ETPF and one defines
\begin{equation}
\bx_j^n = \sum_{i=1}^N \bx_i^{\rm f} \widehat{d}_{ij}
\end{equation}
even for finite particles numbers. Of course, by its very construction, the ETPF underestimates 
the posterior covariance. However, there are corrections available that lead to second-order
accurate implementations \citep{deWiljes17a}. See Section \ref{sec:second-order exact filters} 
for more details.

Following previously introduced notations, localization can now be implemented into the
ETPF as follows. For each grid point $k$, we extract the values of the forecast particle 
$\bx_i^{\rm f}$  at that grid point and denote them by $\bx_i^k$. 
Using the observations local to this grid point, we calculate localized importance weights $w_i^k$ for $\bx_i^k$. 
Then (\ref{ETPF_cost}) gives rise to a localized transformation matrix 
\begin{equation} \label{LETPF_cost}
\widehat{\bf D}^k = \arg \min_{{\bf D}\in {\cal D}^k} 
\sum_{i,j=1}^N d_{ij} \|\bx_i^k - \bx_j^k\|^2
\end{equation}
at grid point $k$ with the set $\mathcal{D}^k$ defined by
\begin{equation}
\mathcal{D}^k =\left \{ {\bf D} \in \mathbb{R}_+^{N\times N}: \,\sum_{i=1}^N d_{ij} = 1,
\sum_{j=1}^N d_{ij} =w_i^k N \right\}.
\end{equation}
Note that the transport cost (distance) $t_{ij} = \|\bx_i^k - \bx_j^k\|^2$ can be replaced by any
other localized cost function. See \cite{reich15} for more details.  
The transport problem (\ref{LETPF_cost}) at each grid point can be computationally
expensive.  Less expensive approximations, such as the Sinkhorn approximation, and their
implementation into the localized ETPF (LETPF) are discussed in \cite{deWiljes17a}.
\cite{Farchi2018} have extended this algorithm to block weighting, similar to their extension 
of the Local Particle Filter.

The latter authors also defined a local transform particle filter in state space. 
This involves a transformation, at each grid point, from prior to posterior 
particles by a transformation, which essentially becomes an anamorphosis step. 
The prior and posterior probability densities need to be known as continuous densities,
and \cite{Farchi2018} use kernel density estimation with the particles as basis.
The interesting suggestion is that since the transformation is deterministic and 
expected to be smooth over the space coordinates, no specific smoothing is needed
after the transformation.
We refer to their paper for details on this methodology.

\subsection{Space-Time Particle Filters}

The idea to run a particle filter over the spatial domain was introduced by \cite{VanLeeuwen2009},
and the first algorithm, the Location Bootstrap Filter,  was published by \cite{Briggs13}.
The Space-Time Particle Filter by \cite{Beskos17}  improves on this algorithm
by removing the jitter step, as explained below.
In the following we assume observations at every grid point, but the algorithms can easily be 
adapted to other observation networks.

The Location Particle Filter of \cite{Briggs13} runs as follows. 
The grid points are ordered $1,...,L$, such that points $l$ and $l+1$ are neighbouring grid points for each $l \in {1,...,L}$.
In each grid point $l$ we have a sample $\bx_{i,l}$ for $i \in {1,...,N}$, and $l$ denotes 
the grid point number.
We start the spatial particle filter at location $l=1$ by calculating the weight $p(\by_1 | \bx_{i,1})$ (where the time index is suppressed) 
for each prior particle $i$, and perform resampling using these weights over the whole spatial domain.
This means that the resampled particles are now samples of $p(\bx^{1:L} | \by^1)$. 
A small amount of jitter is added to avoid identical particles. The choice of this jitter density is again not clear for geophysical applications, 
more research is needed on this issue.

Then, the algorithm moves to the next grid point, calculates the weights $p(\by_2 | \bx_{i,2})$, and resamples the 
full state particles using this weight, generating samples from $p(\bx_{1:L} | \by_1,\by_2)$.
Again some jitter is needed to avoid ensemble collapse, and the algorithm moves to the next grid point, until
all grid points are treated this way. Algorithm \ref{alg:LoPF} describes the computational steps.

\begin{algorithm}[!ht]
	\caption{Location Particle Filter \label{alg:LoPF}}
	\begin{algorithmic}[0]
	        \For {Each grid point $j$, and local grid points $k$}
		        \For {$i=1,..,N$}
			    \State $ w_i \gets p(\by_k|\bx_{i,k})$ 
			\EndFor
			\State $ \bw \gets \bw/\bw^T\mathbf{1}$
			\State Resample 
			\State Define jitter covariance $\bS $
		        \For {$i=1,..,N}$
			   \State $\bbeta \sim N(0,\bS)$
	                    \State $ \bx_{i,j} \gets \bx_{i,j} + \bbeta $
	         	\EndFor
		\EndFor
	\end{algorithmic}
\end{algorithm}

Note that the algorithm does not suffer from artificial sharp gradients because all resampled particles are global particles, but the algorithm will be very sensitive to the choice of the jitter density used after updating
the ensemble in each grid point. Furthermore, when prior and posterior are very different, the algorithm will perform poorly, and
\cite{Briggs13} propose a smoother variant that employs copulas for numerical efficiency. We will not discuss that
variant here.

\cite{Beskos17} introduce the Space-Time Particle Filter. Instead of using a jitter density to avoid identical particles
they exploit the spatial transition density $p(\bx^{n}_l|\bx^{n,1}_{l-1},\bx^{n-1}_{1:L})$, in which $n$ is the time index 
and $l$ the spatial index.
(In fact, \cite{Beskos17} allow for a proposal density, but we will explain the algorithm with using the prior
spatial pdf as proposal.)
So they exploit the pdf of the state at time $n$ and grid point $l$, $\bx^{n}_l$, conditioned on all
previous grid points $\bx^{n}_{1:l-1}$ at the same time $n$, and conditioned on all grid points at time $n-1$,
denoted $\bx^{n-1}_{1:L}$.
They do this by introducing a set of $M$  local particles $j$, for each global particle $i$, with $i \in {1,...,N}$. 

For each of the global particles $i$ they run the following algorithm over the whole grid:
\begin{itemize}
\item[1]Starting from location $l=1$ the $M$ local particle filters grow in dimension when moving over the grid towards
the final position $L$. 
At the first grid point the prior particles at that grid point are used, weighted with the local likelihood $p(\by_1 | \bx_1)$
and resampled. Let us call these particles $\hat{\bx}_{j,1}$, in which
$j$ is the index of the local particle, and $1$ is the index of the grid point.
\item[2]The mean $\bar{w}^1$ of the unnormalised weights  is calculated.
\item[3]For the next grid point each of these $M$ resampled particles are propagated to that grid point by drawing from $p(\bx_2 | \hat{\bx}_{j,1}, \bx_{j,1:L}^{n-1})$.
Since each of the $M$ particles is drawn independently they will differ and no jittering is needed.
\item[4]Then the unnormalised weights $p(\by_2 | \bx_2)$ are calculated, and their mean $\bar{w}^2$, followed by a resampling step.
\item[5]This process is repeated until $l=L$, so until the whole space is covered.
\item[6]Finally, the total weight $w_1 = \prod_{l=1}^L \bar{w}^l$ is calculated, which is the unnormalised weight
of the 1st global particle.
\end{itemize}
Algorithm \ref{alg:STPF} summarises the scheme.

\begin{algorithm}[!ht]
	\caption{Space-Time Particle Filter \label{alg:STPF}}
	\begin{algorithmic}[0]
	     \For {$i=1,..,N$}
	        \For {Each grid point $j$, and local grid points $k$}
		        \For {$m=1,..,M$}
		            \State $ \bx_{m,j}^{n} \sim p(\bx^n_j|\bx^{n}_{1:l-1},\bx^{n-1}_{1:L}) $
			    \State $ \tilde{w}_m \gets p(\by_k|\bx_{i,k})$ 
			\EndFor
			\State $\bar{w}_{i,j} \gets \frac{1}{M} \sum_{m=1}^M \tilde{w}_m $
	            \EndFor
		 \State $ w_i \gets \prod_{j=1}^L \bar{w}_{i,j} $
                \EndFor
                \State $ \bw \gets \bw/\bw^T\mathbf{1}$
                \State Resample
	\end{algorithmic}
\end{algorithm}

This procedure is followed $N$ times for each global particle $i$ independently.
These global particles are then resampled according to the weight $G_i$
It is still possible that this filter is degenerate, see \cite{Beskos17} for details and potential solutions.

The importance of this filter lies in the fact that there is a formal proof that it converges to
the correct posterior for an increasing number of particles, unlike any of the other algorithms discussed.
Furthermore, the authors show that degeneracy can be avoided if the number of particles grows as 
the square of the dimension of the system, indeed much faster convergence than e.g. the optimal proposal density.

\subsection{Discussion}

Following into the footsteps of Ensemble Kalman Filters, exploring localisation in particle filters is a rapidly growing field. 
But localisation in particle filters is not trivial as there is no automatic smoothing via smoothed sample
covariances as in Ensemble Kalman filters. Most local particle filters impose explicit spatial smoothing,
which can affect delicate balances in the system. Worth mentioning in this context is the localisation
introduced by \cite{Robert17b}, who process observations sequentially in their hybrid Ensemble Kalman
Filter-Particle Filter approach such that the second-order properties of the particle-filter part remain correct. This method is discussed in the next chapter.
The Ensemble Transform Particle Filter and 
the Localized Adaptive Particle Filter come closest to the Ensemble Kalman Filter by using a 
linear transportation matrix to transforms the 
prior ensemble into a posterior ensemble, and this matrix can be made 
smoothly varying with space. 
All of these smoothing operations rely on forming linear combinations 
of particles, so can potentially harm nonlinear 
balances in the model. Furthermore, it should be noted that the 
smoothing operation does not necessarily follow
Bayes Theorem, so it might result in an extra 
approximation of the true posterior pdf. When the ensemble size is
small this approximation might be negligible compared to the 
Monte-Carlo noise from the finite ensemble size, however.

The Location Particle Filter and the Space-Time Particle Filter avoid this smoothing and rely on statistical 
connections between different grid points. The former does this via the prior pdf, defined by the prior particles.
When the number of particles is low this pdf is estimated rather poorly.
Furthermore, the method needs jittering of the global particles to avoid 
ensemble collapse after every resampling step after each new observation is assimilated. This jittering pdf 
can be chosen arbitrarily, for instance a smooth Gaussian, but it does violate Bayes Theorem. As mentioned
above, this error might be negligible when the ensemble size is small. 
The latter method explores the transition density over space and time, leading to consistent 
estimates of the spatial relations between grid points.
Another potential issue of both methods is that if the spatial field is two or higher dimensional, as in geoscience applications,
it is unclear how to order the grid points, and potentially large jumps might be created between neighbouring 
grid points that are treated as far apart by the algorithm. This needs further investigation.

\section{Hybrids between Particle Filters and Ensemble Kalman Filters}
\label{sec:Hybrid}
As mentioned in the previous section, there are two issues with localisation. Firstly, particle filters that employ resampling
need to ensure smooth updates in space so that the newly formed global particles do not encounter strong adjustments
to physical balances due to artificial gradients from glueing particles together. Present-day localised particle schemes concentrate on this issue.

Secondly, the localisation area cannot contain too many
independent observations, and as a rule of thumb 10 independent observations is often too many, to avoid weight
collapse. As mentioned, this demand can be in strong contrast with physical considerations of appropriate length scales.
This is one of the main reasons to consider hybrids between particle filters and ensemble Kalman filters within 
a localisation scheme. In the following several recent hybrid methods are presented.

\subsection{Adaptive Gaussian Mixture Filter}

A bridging formulation allows to smoothly transition between an ensemble Kalman filter and a particle filter analysis update. One such formulation is the adaptive Gaussian mixture filter \citep{Stordal11a}.

In a Gaussian mixture filter, the distribution is approximated by a combination of normal distributions centered at the values of the particles. Thus we have
\begin{equation}
p(\bx^n) = \sum_{i=1}^{N} w_i N\left(\bx_i^f, \hat{\bP}^f \right)
\label{GaussianMixture}
\end{equation}
where $N(\bx_i^f, \hat{\bP}^f)$ is a Gaussian Kernel with mean $\bx_i^n$ and covariance $\hat{\bP}^f$. This covariance is initialized from the sample covariance matrix $\bP^f$ of the ensemble by multiplying with a so-called bandwidth parameter $0<h\leq1$ such that
\begin{equation}
\hat{\bP}^f = h^2 \bP^f .
\end{equation}

At the analysis time, the filter computes a two-step update: In the first step we update the ensemble members and the covariance matrix according to the Kalman filter equations given by
\begin{eqnarray}
\bX^n &=& \bX^f + \hat{\bK}^n \left(\by^n\mathbf{1}^T - \bH \bX^f\right)\\
\hat{\bK}^n &=& \hat{\bP}^f \bH^T \left(\bH \hat{\bP}^f \bH^T + \bR^n\right)^{-1}
\end{eqnarray}
and
\begin{eqnarray}
\bP^n &=& \left(\bI - \hat{\bK}^n \bH\right)\hat{\bP}^f.
\end{eqnarray}
Note that this is just a short-hand notation for updating each centre fo the prior Gaussians.
For computational efficiency the analysis equations in the (adaptive) Gaussian mixture filter \citep{Hoteit08a, Stordal11a} were proposed to use a factorized covariance matrix in the form $\hat{\bP}^f = \bL \bU \bL^T$ as can be obtained from a singular value decomposition of the ensemble perturbation matrix and used, e.g. in the SEIK filter \citep{Pham01a} and error-subspace transform Kalman filter \citep[ESTKF,][]{Nerger12a}. However, the particular form of the Kalman filter update equations is not crucial here.

In the second step we update the weights of the particles according to
\begin{equation}
w_i^n \approx w_{i}^{n-1} N_{\by^n|\bx^f}\left(\bH \bx_i^f, \bR^n\right)
\end{equation}
in which $\bR^n = \bR + \bH \hat{\bP}^f \bH^T$, and then normalise these so that the sum of the weights is one. 

The bridging is now done by interpolating the analysis weight with a uniform weight $N^{-1}$ as
\begin{equation}
w_i^{(\alpha)} = \alpha w_i + (1-\alpha)N^{-1},
\end{equation}
where $\alpha$ is the bridging parameter. We obtain a transition between the ensemble Kalman filter and the particle filter by varying both $\alpha$ and $h$. For $\alpha=0$ and $h=1$ we obtain the uniform weights of the ensemble Kalman filter, while for $\alpha=1$ and $h=0$ we obtain the particle filter weights. \cite{Stordal11a} proposed to adaptively estimate an optimal value of $\alpha$ by setting $\alpha = N^{-1} \hat{N}_{eff}$ where $\hat{N}_{eff} = (\sum_i w_i^2)^{-1}$ is the effective sample size.

The update formulation of the adaptive Gaussian mixture filter reduces the risk of ensemble degeneracy, but cannot fully avoid it. To this end, we can combine the filter with a resampling step as in other particle filters. 

\subsection{Ensemble Kalman Particle Filter}

The Ensemble Kalman Particle Filter  of \cite {Frei13} is a hybrid
EnKF-PF. It is based on tempering in just two steps, splitting the likelihood
into two factors
\begin{equation}
p(\bx^n|\by^n) = p(\bx^n|\by^n)^\alpha \;p(\bx^n|\by^n)^{1-\alpha}
\end{equation}
with $\alpha \in (0,1)$. 
In the first step the Stochastic Ensemble Kalman filter of \cite{Burgers98a} is applied, 
and in the second step a particle filter. When the parameter $\alpha$ is close to 0
the scheme is like a full particle filter, while for $\alpha$ close to 1
it is essentially the ensemble Kalman filter.
Figure \ref{fig:EnsembleKalman} illustrates the idea.

\begin{figure}[h]
	\centering\includegraphics[width=1\linewidth]{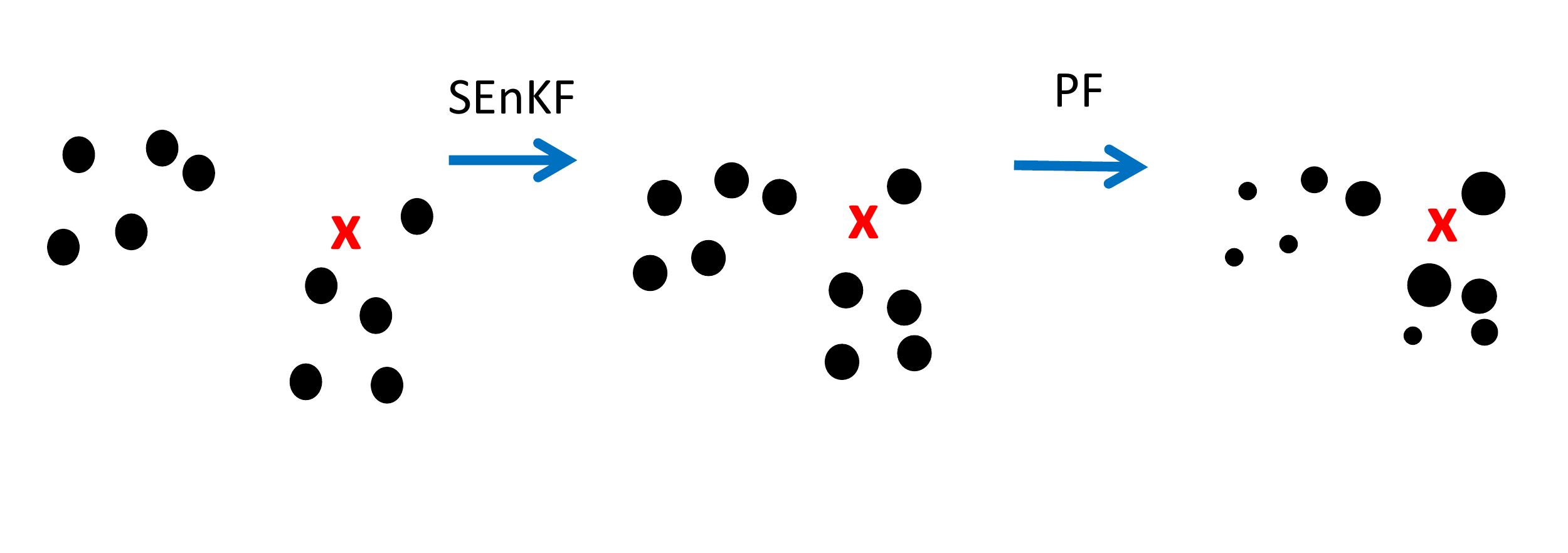}   
	\caption{The Ensemble Kalman Particle Filter. First a Stochastic EnKF is performed, followed by a standard
	Particle Filter.} 
	\label{fig:EnsembleKalman}
\end{figure}    

Two problems with a direct application of the above scheme are identified by \cite{Frei13}:
the particle filter weights are influenced by the random modelled observations in the Stochastic EnKF (SEnKF), and
the resampling step in the particle filter will lead to identical particles.
To avoid both, the algorithm is modified as follows. Firstly, assuming a Gaussian likelihood, the SEnKF particles can be written as:
\begin{equation}
\bx_i ^{SEnKF} = \bx_i + \bK_{\alpha}(\by-\bH \bx_i - \bepsilon_i)
\end{equation}
with $\bepsilon_i \sim N(0,\bR/\alpha)$ and
$\bK_{\alpha}$ is the normal gain, but with $\bR$ divided by $\alpha$. Thus, the particles
can be seen as draws from 
\begin{equation}
\bx_i^{SEnKF} \sim N(\bnu_i,\bP^{EnKF}) 
\end{equation}
in which
\begin{equation}
\bnu_i = \bx_i + \bK_{\alpha} (\by-\bH \bx_i)
\end{equation}
and 
\begin{equation}
\bP^{SEnKF} =\frac{1}{\alpha} \bK_{\alpha}\bR \bK_{\alpha}^T .
\end{equation}
Hence the SEnKF posterior can be written as:
\begin{equation}
p(\bx|\by)^{SEnKF} = \frac{1}{N}\sum_{i=1}^N  N(\bnu_i,\bP^{SEnKF}) 
\end{equation}
Instead of performing the standard SEnKF sampling from this density we delay that sampling and perform the multiplication 
with the second likelihood $p(\by|\bx)^{1-\alpha}$ analytically.
This is easy because the EnKF posterior is a Gaussian mixture and the likelihood is a Gaussian, 
so the full posterior is a Gaussian mixture too. 
This leads to a full posterior
\begin{equation}
\label{eq:enkpf-mixt}
\sum_{i=1}^N \gamma_i N(\bmu_i,\bP^{PF})
\end{equation}
in which 
\begin{eqnarray}
\bmu_i &  = & \bnu_i + \hat{\bK} (\by-\bH\bnu_i)  \\
\gamma_i & = & N\left(\by-\bH\bnu_i, \bH \bP^{SEnKF}\bH^T + \bR/(1-\alpha)\right)  \\
\bP^{PF} & = & (\bI-\hat{\bK}H)\bP^{SEnKF}
\end{eqnarray}
where
\begin{equation}
\hat{\bK} = \bP^{SEnKF}\bH^T \left(\bH \bP^{SEnKF} \bH^T + \bR/(1-\alpha) \right)^{-1}
\end{equation}
Note that the normalisation constants in $\gamma_i$ do not have to be calculated as we 
know that they should fulfil $\sum_i \gamma_i=1$.

The way to sample the particles now becomes a two step procedure.
First draw $N$ samples from the distribution of the mixture coefficients
$\gamma_i$ and then draw from the selected Gaussian mixture components:
\begin{equation}
\bx_i^{EKPF} = \bmu_{k_i} + \bxi_{i}
\end{equation}
in which $k_i$ denotes the resampled particle index $i$ and
$\bxi_{i} \sim N(0,\bP^{PF})$. The variables $\bxi_i$ can again be
generated in two steps by
\begin{equation}
\bxi_i= (\bI - \hat{\bK} H^T) \bK_{\alpha} \bepsilon_{i,1} + \hat{\bK} \bepsilon_{i,2}
\end{equation}
where $\bepsilon_{1.i}$ and  $\bepsilon_{i,2}$ are independent draws from
$N(0,\bR/\alpha)$ and $N(0,\bR/(1-\alpha))$, respectively.

The scheme is very closely related to a Gaussian mixture model, as the EnKF step forces the 
prior for the particle filter to be a Gaussian mixture.
The strong point of this scheme is that the width of each Gaussian follows naturally from 
the stochastic part of the EnKF, while it is ad hoc in standard Gaussian mixture models.
Furthermore, while the standard Gaussian mixture model uses the observation
covariance matrix $\bR$ this filter uses an inflated $\bH \bP^{SEnKF} \bH^T + \bR/(1-\alpha)$, which will lead
to a better weight distribution. Finally, the starting points of the centres of the prior Gaussians
will be closer the observations, suggesting more uniform weights. The pseudocode of the 
scheme is presented in Algorithm \ref{alg:EKPF}.

\begin{algorithm}[!ht]
	\caption{Ensemble Kalman Particle Filter \label{alg:EKPF}}
	\begin{algorithmic}[0]
	    \State $ \bR_{\alpha} \gets \bR/\alpha $
	    \State $ \bK_{\alpha} \gets \bP \bH^T (\bH \bP \bH^T + \bR_{\alpha})^{-1} $
	    \State $ \bP^{SEnKF} \gets\frac{1}{\alpha} \bK_{\alpha}\bR \bK_{\alpha}^T $
	    \State $ \hat{\bK} \gets \bP^{SEnKF}\bH^T \left(\bH \bP^{SEnKF} \bH^T + \bR/(1-\alpha) \right)^{-1} $
	    \For {$i=1,..,N$}
	                \State $ \bepsilon_{i,1} \sim N(0,\bR_{\alpha}) $
	                \State $ \bepsilon_{i,2} \sim N(0,\bR/(1-\alpha) ) $
	                \State $ \bnu_i \gets \bx_i + \bK_{\alpha}(\by-\bH \bx_i - \bepsilon_i) $
	                \State $ \bmu_i  \gets  \bnu_i + \hat{\bK} (\by-\bH\bnu_i) $
	               \State $ \gamma_i \sim N\left(\by-\bH\bnu_i, \bH \bP^{SEnKF}\bH^T + \bR/(1-\alpha)\right) $
	  \EndFor
	  \State $ \bgamma \gets \bgamma  /\bgamma^T \mathbf{1} $
	  \For  {$i=1,..,N$}
	             \State $ k_i \sim MultiNomial(\bgamma) $
	             \State $ \bxi_i \gets (\bI - \hat{\bK} H^T) \bK_{\alpha} \bepsilon_{i,1} + \hat{\bK} \bepsilon_{i,2} $
	             \State $ \bx_i^{EKPF} \gets \bmu_{k_i} + \bxi_i $
                \EndFor
                \State Resample
	\end{algorithmic}
\end{algorithm}

In an extension of the scheme, \cite{Frei13} suggest to form a tempering scheme,
alternatively using the ensemble Kalman filter and the particle filter.
The resampling step of the particle filter is not problematic in this case as the Kalman filter 
will diversify identical particles in each next iteration.
The paper also discusses approximate schemes for non-Gaussian observation errors 
and nonlinear observation operators.

In \cite {Robert17a}, a variant of this method has been introduced which 
is based on the LETKF instead  of the stochastic
variant and in which the update is in ensemble space:
\begin{equation}
  \bX^{PI} = \bX^f \bW
\end{equation}
where the column sums of $\bW$ equal 1. The matrix $\bW$ can be split into
\begin{equation}
  \bW= \bW^\mu \bW^\alpha + \bW^\xi
\end{equation}
where $\bW^\mu$ corresponds to computing the centers $\bmu_i$, $\bW^\alpha$
to the resampling and $\bW^\xi$ to the added noise $\bxi_i$. In the
transform variant $\bW^\xi$ is deterministic and chosen such that
the sample covariance of $\bX^{PI}$ is equals the covariance of the
Gaussian mixture  \eqref{eq:enkpf-mixt}. It thus belongs also to the
class of second-order exact filters discussed in the next section.

\cite{Robert17a} apply a localized transform Ensemble Kalman Particle
Filter in the KENDA (Kilometer-Scale Ensemble Data Assimilation) system with
a setup similar to the one used operationally by MeteoSwiss. This 
system computes the weight matrices $\bW$ only on a coarse grid and
then interpolates these matrices to the original grid. Therefore the
discontinuities introduced by resampling are smoothed out, but in a
way that is possibly optimal for the EnKF and not for the EnKPF.
In \cite{Robert17b} a different localization method for the EnKPF
was developed which proceeds by sequentially assimilating observations
$y^k$, limiting the state components influenced by $y^k$ to a subset.
It smoothes out the discontinuities that occur when a resampled particle 
in the region influenced by $y^k$ is connected to a background particle outside
of this region. The smoothing is  done in such a way that the 
second-order properties of the smoothed particle remain correct.

\subsection{Second-order exact filters} \label{sec:second-order exact filters}

A second-order exact filter ensures that the posterior ensemble mean and ensemble covariance matrix are equal to those obtained from the particle filter weights. Thus, the requirement for the mean of the analysis ensemble is 
\begin{equation}
\label{eq:sndorderX}
\overline{\bx}^n = \frac{1}{N} \sum_{i=1}^{N} \bx^n_i = \sum_{i=1}^{N} w_i \bx^f_i
\end{equation}
where the superscript $f$ denotes the forecasted state vector. Likewise, the posterior ensemble covariance matrix is required to fulfil
\begin{eqnarray}
\bP^a & = & \frac{1}{N} \sum_{i=1}^{N} \left(\bx^n_i - \overline{\bx}^n\right) \left(\bx^n_i - \overline{\bx}^n\right)^T\\
\label{eq:sndorderP}
 & = & \sum_{i=1}^{N} w_i \left(\bx^f_i - \overline{\bx}^n\right) \left(\bx^f_i - \overline{\bx}^n\right)^T .
\end{eqnarray}

\subsubsection{Merging Particle Filter}

The merging particle filter by \cite{Nakano2007} explores the sampling aspect of the resampling step.
The method draws a set of $q$ ensembles each of size $N$ from
the weighted prior ensemble at the resampling step.
Then these sets are merged via a weighted
average to obtain a new set of particles that has the correct mean and covariance but is more robust than
the standard particle filter.  Define $\bx_{i,j}$ as
ensemble member $i$ in ensemble $j$. The new merged ensemble members are generated
via
\begin{equation}
\bx_i^a = \sum_{j=1}^q \alpha_j \bx_{i,j} .
\end{equation}
To ensure that the new ensemble has the correct mean and covariance, the coefficients $\alpha_j$ have to be real and need to fulfil the two conditions 
\begin{equation}
\sum_{j=1}^q \alpha_j=1;  \quad \sum_{j=1}^q \alpha_j^2=1 ,
\end{equation}

When $q>3$ there is no unique solution for the $\alpha$'s, while for $q=3$ one finds:
\begin{eqnarray}
\alpha_1 & = & \frac{3}{4}  \nonumber \\ 
\alpha_2 & = & \frac{\sqrt{13}+1}{8}   \nonumber \\
\alpha_3 & = & -\frac{\sqrt{13}-1}{8}
\end{eqnarray}
We can make the weights space-dependent in high-dimensional systems and 
since the new particles are merged previous particles the resulting global particles are expected to be smooth.
The scheme is depicted in Algorithm \ref{alg:MPF}.

\begin{algorithm}[!ht]
	\caption{Merging Particle Filter \label{alg:MPF}}
	\begin{algorithmic}[0]
		   \For {$i=1,..,N$}
			    \State $ w_i \gets p(\by_k|\bx_{i})$ 
			\EndFor
  		   \State $ \bw \gets \bw/\bw^T\mathbf{1}$
	         \State $(\bX^a_1,..., \bX^a_q) \gets $ $q$ times resampled prior ensemble 
	         \State Find $\alpha_i$ such that $\sum_i \alpha_i=1$ and $\sum_i \alpha_i^2=1$  
	         \State $ \bX^a \gets \sum \alpha_i \bX^a_i$ 
	\end{algorithmic}
\end{algorithm}

\subsubsection{Nonlinear Ensemble Transform Filter NETF}

A simple formulation of a second-order exact filter can be obtained by using Eq. (\ref{eq:sndorderX}) to compute the mean of the posterior ensemble \citep{Xiong06a, Toedter15a}. For the associated ensemble perturbations, 
we can derive from Eq. (\ref{eq:sndorderP}) with $\mathbf{w} = (w_1, \ldots, w_N)^T$ and $\mathbf{W} = \mathrm{diag}(\mathbf{w})$ that
\begin{equation}
\bP^a = \bX^f \left(\mathbf{W} - \mathbf{w}\mathbf{w}^T\right) (\bX^f)^T\ .
\end{equation}
Posterior ensemble perturbations can now be obtained by factorizing $\mathbf{A} = \mathbf{W} -\mathbf{ww}^T$, e.g. by a singular value decomposition as $\mathbf{A} = \mathbf{V\Lambda V}^T$. This leads to $\mathbf{A}^{1/2} = \mathbf{V\Lambda}^{1/2}\mathbf{V}^T$ and posterior perturbations are then given by
\begin{equation}
\bX'^n = \sqrt{N} \bX^f \mathbf{V\Lambda}^{1/2}\mathbf{V}^T.
\end{equation}
Finally, the full posterior particles are given by
\begin{equation}
\label{eq:netf_update}
\bx_i^n = \bX^f \left(\mathbf{w}\mathbf{1}^T + \sqrt{N}\mathbf{V\Lambda}^{1/2}\mathbf{V}^T\right)_i.
\end{equation}
The computations of this filter are very similar to those in ensemble square-root Kalman filters like the ETKF \citep{Hunt07a} or ESTKF \citep{Nerger12a}. As such, we can can also localize the filter in the same way. The localized NETF has been successfully applied to a high-dimensional geophysical system based on primitive equations in \cite{Toedter16a}. In addition, the filter can be easily extended to a smoother by applying the filter transform matrix (the term in parenthesis in Eq.\ \ref{eq:netf_update}) to previous analysis times \citep{Kirchgessner17a}.
The scheme is depicted in Algorithm \ref{alg:NETF}.

\begin{algorithm}[!ht]
	\caption{NETF  \label{alg:NETF}}
	\begin{algorithmic}[0]
		\For {$i=1,..,N$}
			 \State $ w_i \gets p(\by_k|\bx_{i})$ 
	   \EndFor
		\State $ \bw \gets \bw/\bw^T\mathbf{1}$
		\State $\bA \gets diag (\bw) - \bw\bw^T$ 
		\State $\bV\Lambda \bV^T \gets \bA$ 
		\State $\bT \gets \sqrt{N}\bV \Lambda^{1/2} \bV^T$ 
		\State $\bT \gets \bT + \bw$ 
		\State $\bX^a \gets  \bX^f \bT$ 
	\end{algorithmic}
\end{algorithm}

\subsubsection{Nonlinear Ensemble Adjustment Filter}

There is also a stochastic variant of the previous algorithm \citep{Lei11a}, which is motivated from the Stochastic Ensemble Kalman filter \citep{Burgers98a, Houtekamer98a}. In this filter, we generate a set of perturbed model observations 
\begin{equation}
\by_i = \bH(\bx_i) + \bepsilon_i,\;\;\; i=1,\ldots, N,
\end{equation}
which represents the observation probability distribution. We now obtain an analysis mean of each particle analogously to Eq. (\ref{eq:sndorderX}) by
\begin{equation}
\label{eq:NLEAFX}
\overline{\bx}^n(\by_k) = \sum_{i=1}^{N} w_i(\by_k) x^f_i\end{equation}
where each weight $w_i(\by_k)$ is computed from the likelihood of the perturbed measured ensemble member $\bH(\bx_i)$ . When we now define
\begin{equation}
\hat{\bP}^a(\by_k)  =  \sum_{i=1}^{N} w_i(\by_k) \left(\bx^f_i - \overline{\bx}^n(\by_k)\right) \left(\bx^f_i - \overline{\bx}^n(\by_k)\right)^T
\end{equation}
we obtain the posterior ensemble members as
\begin{equation}
\bx^n_k = \overline{\bx}^n + (\bP^a)^{1/2} \hat{\bP}^a(\by_k)^{-1/2} (\bx^f_k - \overline{\bx}^n(\by_k))
\end{equation}
where $\overline{\bx}^n$ is given by Eq. (\ref{eq:sndorderX}) and $P^a$ is given by Eq. (\ref{eq:sndorderP}). 
This update equation only yields the correct first and second moments of the posterior distribution in the limit of a large ensemble.

\subsubsection{Second-order exact ETPF}

Also the ETPF (see Sec. \ref{sec:LETPF}) can be formulated to be second-order accurate \citep{deWiljes17a}. For this, we approximate
\begin{equation}
\label{eq:sndorderA}
\mathbf{A} = \mathbf{W} -\mathbf{ww}^T \approx \frac{1}{N}\left(\widehat{\mathbf{D}} -\mathbf{w1}^T\right)\left(\widehat{\mathbf{D}} -\mathbf{w1}^T\right)^T
\end{equation}
where the matrix $\widehat{\mathbf{D}}$ is obtained through (\ref{ETPF_cost}). 
To ensure the second-order accuracy, we introduce a correction term such that
\begin{equation}
\widetilde{\mathbf{D}} = \widehat{\mathbf{D}} + \mathbf{\Delta}
\end{equation}
with $\mathbf{\Delta}$ being a symmetric $N \times N$ matrix. Using $\widetilde{\mathbf{D}}$ in Eq. (\ref{eq:sndorderA}) and requiring that the result is equal to $\mathbf{A}$ leads to the condition
\begin{eqnarray}
N(\mathbf{W} -\mathbf{ww}^T) - (\widehat{\mathbf{D}}-\mathbf{W1}^T)(\widehat{\mathbf{D}}-\mathbf{W1}^T)^T\quad\quad\quad\\
=(\widehat{\mathbf{D}}-\mathbf{W1}^T)\mathbf{\Delta} + \mathbf{\Delta}(\widehat{\mathbf{D}}-\mathbf{W1}^T)^T + \mathbf{\Delta}\mathbf{\Delta},
\end{eqnarray}
which is a quadratic equation in $\mathbf{\Delta}$ in the form of a continuous-time algebraic 
Riccati equation and there are known solution methods for this type of equation \citep[see, e.g.,][]{deWiljes17a}. Note that $\widetilde{\bf D}$ still satisfies (\ref{marginals}). However, 
$\widetilde{d}_{ij} \ge 0$ does not hold anymore, in general.

\subsection{Hybrid LETPF-LETKF}
The hybrid LETPF-LETKF is also based on the simple idea of splitting
the likelihood function into two factors at each grid point $k$, i.e.
\begin{equation}
p(\bx^k|\by^{(k)}) = p(\bx^k|\by^{(k)})^{1-\alpha}\ p(\bx^k|\by^{(k)})^{\alpha}
\end{equation}
with $\alpha \in (0,1)$, but now the particle filter is employed first, followed by
the ensemble Kalman filter.  This is similar to tempering in just two steps.
When the likelihood is Gaussian the posterior is expected to be more Gaussian than the prior.
Hence it makes sense to use a particle filter in the first step, and
to try to use an EnKF in the second step of the tempering procedure.

If the likelihood is Gaussian with localized error covariance matrix
$\bR^k$, then the factorization is equivalent to scaling this matrix  by $1/\alpha$
and $1/(1-\alpha)$, respectively. Hence, one can, for example, first apply an LETPF
to the forecast particles $\bx_i^{\rm f}$ with inflated covariance matrix $\bR^k/\alpha$
in order to obtain new particle values
\begin{equation}
\tilde \bx_i^k = \sum_{j=1}^N d_{ij}^k(\alpha) \bx_i^k
\end{equation}
at each grid point $k$. One then applies the LETKF to these intermediate particles $\tilde \bx_i$
with inflated covariance matrix $\bR^k/(1-\alpha)$. The choice of $\alpha$ is, of course, 
crucial. Numerical experiments indicate \citep{CRR15} that $\alpha>0$ can lead to 
substantial improvements over a purely LETKF-based implementation and that the choice of $\alpha$
can be based on the effective sample size of the associated LETPF. However, more
refined selection criteria for the parameter $\alpha$ are needed to make the hybrid 
LETPF-LETKF method widely applicable. 

\subsection{Hybrid EnVar PF}
Based on the localized adaptive particle filter (LAPF) described in 
Section \ref{sec:LAPF}, a hybrid particle filter based ensemble
variational data assimilation system (PfVar) can also be constructed.
The idea is to replace the LETKF-based ensemble in an EnVar by an 
LAPF-based ensemble. 

We briefly discuss a practical numerical weather prediction example here.
Following \cite{Buehner2013},  the operational EnVAR system of DWD for the ICON 
model with 13km global resolution and 6.5km resolution of its two-way nested 
area over Europe is using the ensemble of the global 40 member LETKF for its 
dynamic covariance matrix with a ratio of 70:30 towards the classical NMC 
based covariance matrix of the three-dimensional variational data 
assimilation system with 3h cycling interval. 
The LETKF ensemble is replaced by the LAPF ensemble, where the quality control of the
variational high-resolution run is used for the ensemble data assimilation
system under consideration. In the current system, no recentering of the
ensemble with respect to the variational mean estimator is carried out, leading
to a form of weak coupling of the systems.  

In a quasi-operational setup (without a high-resolution nest), the hybrid PfVAR is running 
stably for a period of one month. The observation minus background statistics 
show very promising behaviour in several case studies which are under 
investigation at DWD \citep{walter17_1}. In the current 
state of tuning, the forecast quality of the PfVAR seems comparable to the 
forecasts based on the LETKF-based EnVAR. These new results studied in 
combination with \cite{Robert17a} show that today's particle 
filters are approaching 
the quality of state-of-the-art operational ensemble data assimilation 
systems and are already becoming important tools on all scales of
NWP. 

\subsection{Discussion}
Hybrid particle-ensemble Kalman filter schemes, especially when implemented adaptively, can avoid 
weight collapse in the particle filter part of the hybrid in any situation. The price paid is that not all
information from the observations is extracted when the posterior pdf is severely non-Gaussian, but
in many situations this is not the dominant source of error. The reason why these schemes are competitive
is that they do take into account some non-Gaussianity via the particle filter, while the particle filter
alone is very inefficient compared to the Ensemble Kalman Filter when the posterior is actually close to a 
Gaussian. So the objective is not necessarily to make the $\alpha$ as small as possible, but indeed
to find an optimal $\alpha$ to ensure that the Ensemble Kalman Filter is used whenever we can. 
The same is true for the bridging parameter in the Adaptive Gaussian Mixture Filter. 

The second-order exact filters are hybrids of a different kind, focussing on obtaining the posterior mean and the covariance
correct given the limited prior ensemble.
These methods are expected to be quite competitive
to the hybrid filters discussed above, and the relative performance will depend strongly on the measure used to
define what is best. For instance, RMSE are expected to be better for the second-order exact filters, while
full ensemble measures like rank histograms and continuous ranked probability scores might benefit
from the hybrid schemes. 

One question that emerges when comparing the Ensemble Kalman Particle Filter and the LETPF-LETKF hybrid
is what should one use first, the particle filter or the ensemble Kalman filter. Different experimental results
seem to indicate that both orderings can be superior to the other. The PF-first methods have the advantage of
a theoretical justification via a two-step tempering interpretation in which the particle filter step makes the 
prior for the EnKF much more Gaussian. Applying the EnKF first will bring the particles closer
to the observations, leading to better weight balance in the particle filter. At this moment it is unclear which order
is best when, much more research is needed.

\section{Conclusions and discussion}

The largest issue of standard particle filters was until recently their degeneracy 
in high-dimensional settings: when the number of independent observations is large and
the number of particles is limited (of order 10-1000 for geophysical applications), one particle
gets weight one, and all others get weight zero.

Two developments have revived the interest in particle filters: efficient proposal densities and localisation,
while hybrids with Ensemble Kalman Filters and recently transportation filters enhance confidence in the usefulness
of particle filters in high-dimensional settings. The new kid on the block are particle flow methods.
Their popularity in the large machine-learning community ensures rapid progress here, too.
It is unclear at this moment how competitive these new ideas will be.
It is clear that developments on particle filters have been very fast,
and the first tests of both localised and hybrid particle-EnKF filters in operational numerical weather prediction
have been performed and show highly encouraging results. 

This paper discussed these new developments and demonstrates that particle filters are useful in even the largest
dimensional geophysical data-assimilation problems and will allow us to make large steps towards fully nonlinear
data assimilation. The emphasis was here on explaining and connecting existing and new ideas, including new understanding of the optimality of the optimal proposal density and equal-weight filters. 

From the presentation it has become clear that the field is too young to provide solid guidance on which method
will be most fruitful for which problem. Given that most data-assimilation practitioners will have an implementation of
a local Ensemble Kalman Filter in some form, localised particle filters seem to be the fastest way to make progress.
However, one has to keep in mind that the resampling step needs smoothing that is more complex than in an Ensemble
Kalman Filter, although exciting new variants like the ETPF and LAPF allow for smooth updates in a very natural way.
Furthermore, with the small ensemble sizes now practical (10-100), more than 10 independent observations in a localisation
area may already lead to filter degeneracy, forcing us to look into methods that limit the weights from below.
This is another ad-hoc procedure that limits information extraction from observations, but it is unclear how severe this
issue is.

Even easier are implementations of hybrid PF-EnKF filters, but it is still unclear what these filters target.
At the moment their value lies in bringing more non-Gaussianity into Ensemble Kalman Filters, but at the same time
ensure that an Ensemble Kalman Filter is used when that is warranted. 

We discussed two main variants that try to avoid localisation because of the issues discussed above:
the equal-weight particle filters and transportation particle filters.
The equal-weight variants, which avoid weight collapse by construction, do not have a complete mathematical foundation yet. 
We know these schemes are biased, but since they
are tailored to high-dimensional problems with small ensemble sizes the bias error might be smaller than the
Monte-Carlo error from the small ensemble size. 
Transportation particle filters still have to demonstrate their full potential in geoscience applications,
but initial experiments with e.g. mapping particle filters on low-to-moderate dimensional systems together with 
the way they are formulated suggest they could become mainstream competitive schemes. 

All in all, huge progress has been made in particle filtering, and 
initial attempts to implement the schemes into full-scale
numerical weather prediction models have succeeded, 
with promising initial results. This shows that particle filters
can no longer be ignored for high-dimensional geoscience applications.

\begin{appendices}

\section{Law of total variance}
The law of total variance is an elementary theorem in statistics and probability. It can be proven as follows.
First we need the Law of total expectation, which reads, using $E_A[B]$ as denoting
the expectation of $B$ under pdf $p(a)$:
\begin{eqnarray}
E_Y[E_{X|Y}[f(X)]] & = &  \int \int f(x) p(x|y) p(y)\; dx\;dy \nonumber \\
& = &   \int_x \int_y f(x) p(x,y) \;dy\;dx \nonumber \\
& = & \int f(x) p(x) \;dx  \nonumber \\
& = &  E_X[f(X)]
\end{eqnarray}
Using this equality on $var_{\bX}[\bX]$ leads to:
\begin{eqnarray}
var_{\bX}[\bX] & = & E_{\bX}[\bX^2] - E_{\bX}^2[\bX]    \nonumber \\
& = & E_Y \left[E_{\bX|Y}[\bX^2]\right] - E_Y^2[E_{\bX|Y}[\bX]]          \nonumber \\
& = & E_Y \left[var_{\bX|Y}[\bX] + E_{\bX|Y}^2[\bX] \right]-  E_Y^2[E_{\bX|Y}[\bX]]         \nonumber \\
& = & E_Y \left[var_{\bX|Y}[\bX]\right] + E_Y\left[E_{\bX|Y}^2[\bX]\right] - E_Y^2[E_{\bX|Y}[\bX]]     \nonumber \\
& = & E_Y\left[var_{\bX|Y}[\bX]\right] + var_Y\left[E_{\bX|Y}[\bX]\right]
\end{eqnarray}
which proves the theorem.

\end{appendices}

\ack PJvL thanks the European Research Council (ERC) for funding the CUNDA project under the European Union Horizon 2020 research and innovation programme. The research of SR has been partially funded by 
Deutsche Forschungsgemeinschaft (DFG) through grant 
CRC 1294 \lq\lq Data Assimilation (project A02)\rq\rq. We thank 3 anonymous reviewers for their critical comments, leading to a much improved paper.


\end{document}